\documentclass[a4paper,11pt]{article}
 \usepackage{jheppub}
\usepackage{amsmath,bm,amssymb,amsthm,mathrsfs,mathtools,multirow}
\usepackage{subfigure,color}
\usepackage{ragged2e}
\justifying\let\raggedright\justifying

\allowdisplaybreaks[4]
\usepackage{slashed}
\usepackage{orcidlink}

\usepackage{tikz}
\usetikzlibrary{backgrounds}
\usetikzlibrary{shapes,matrix,trees}
\usetikzlibrary{arrows.meta}
\usetikzlibrary{positioning}			
\usetikzlibrary{calc,through}			
\usetikzlibrary{decorations.pathreplacing}     
\usepackage{pgffor}                                        
\usetikzlibrary{decorations.pathmorphing}	
\usetikzlibrary{decorations.markings}
\tikzset{
	vector/.style={decorate, decoration={snake}, draw},
	provector/.style={decorate, decoration={snake,amplitude=2.5pt}, draw},
	antivector/.style={decorate, decoration={snake,amplitude=-2.5pt}, draw},
	fermion/.style={draw=black, postaction={decorate},
		decoration={markings,mark=at position .55 with {\arrow[draw=black]{>}}}},
	fermionbar/.style={draw=black, postaction={decorate},
		decoration={markings,mark=at position .55 with {\arrow[draw=black]{<}}}},
	fermionnoarrow/.style={draw=black},
	gluon/.style={decorate, draw=black,
		decoration={coil,amplitude=4pt, segment length=5pt}},
	scalar/.style={dashed,draw=black, postaction={decorate},
		decoration={markings,mark=at position .55 with {\arrow[draw=black]{>}}}},
	scalarbar/.style={dashed,draw=black, postaction={decorate},
		decoration={markings,mark=at position .55 with {\arrow[draw=black]{<}}}},
	scalarnoarrow/.style={dashed,draw=black},
	electron/.style={draw=black, postaction={decorate},
		decoration={markings,mark=at position .55 with {\arrow[draw=black]{>}}}},
	bigvector/.style={decorate, decoration={snake,amplitude=4pt}, draw},
	photon/.style={decorate, draw=black,decoration={snake,amplitude=4pt, segment length=5pt} }
}

\definecolor{ccblue}{rgb}{0.0,0.4,0.8}
\usepackage[]{hyperref}
\hypersetup{  colorlinks=true,
	linkcolor=ccblue,
	urlcolor=ccblue,
	citecolor=ccblue}

\title{Revisiting \texorpdfstring{$\bm{\mu}$-$\bm{e}$}{} conversion in \texorpdfstring{$\bm{R}$}{}-parity violating SUSY}

\author[a,b]{Yu-Qi Xiao\orcidlink{0000-0002-9566-7312}}
\author[a,b]{Xiao-Gang He\orcidlink{0000-0001-7059-6311}}
\author[a,c]{Hong-Yi Niu\orcidlink{0009-0003-1513-540X}}
\author[a,b]{Rong-Rong Zhang\orcidlink{0009-0009-6230-5014}}

\affiliation[a]{State Key Laboratory of Dark Matter Physics, \\
Tsung-Dao Lee Institute and School of Physics and Astronomy, \\
Shanghai Jiao Tong University, Shanghai 201210, China}
\affiliation[b]{Shanghai Key Laboratory for Particle Physics and Cosmology, \\
Key Laboratory for Particle Astrophysics and Cosmology (MOE), School of Physics and Astronomy, Shanghai Jiao Tong University, Shanghai 201210, China}
\affiliation[c]{Zhiyuan College, Shanghai Jiao Tong University, Shanghai 201210, China}

\emailAdd{sjtu7352716@sjtu.edu.cn}
\emailAdd{hexg@sjtu.edu.cn}
\emailAdd{\\shjtu-nhy@sjtu.edu.cn}
\emailAdd{z\_rrong@sjtu.edu.cn}

\abstract{
The $\mu$-$e$ conversion process is one of the most powerful ways to test lepton-flavor-violating (LFV) interactions involving charged leptons. The standard model with massive neutrinos predicts an extremely low rate for $\mu$-$e$ conversion, making this process an excellent probe for testing LFV arising from new physics. Among many theoretical models that can induce LFV, the Supersymmetric model with R-parity violating interactions is one of the most studied for $\mu$-$e$ conversion. 
In this work, we revisit trilinear $R$-parity violating interactions for $\mu$-$e$ conversion, considering renormalization group (RG) running effects from high to low energy scales. The $\mu$-$e$ conversion, $\mu \to e \gamma$, and $\mu \to eee$ experimental data are compared to give upper limits on the relevant 15 combinations of the trilinear $\lambda^{\prime}$ couplings and 6 combinations of the $\lambda$ couplings, certain of which are underexplored in previous studies. We find that RG running effects influence the limits by no more than 30\% in most cases, but can improve constraints by $\sim$80\% in certain combinations, which cannot be neglected. In the near future, COMET and Mu2e are expected to begin data-taking and aim to provide the most stringent constraints on $\mu$-$e$ conversion. These next-generation $\mu$-$e$ experiments have the ability to give much more comprehensive examinations on most trilinear coupling combinations than the $\mu\to e\gamma$ and $\mu\to 3e$ decay experiments. The $\mu$-$e$ experiments will not only deepen our understanding of LFV but also provide a crucial way to examine the underlying new physics contributions.}

\arxivnumber{}

\begin{document}
\maketitle

\section{Introduction\label{intro}}
The $\mu$-$e$ conversion process provides a sensitive probe of charged lepton flavor violation (cLFV). The standard model with massive neutrinos, as observed by neutrino oscillation experiments, can induce this conversion via the one-loop diagram. The predicted conversion rate is negligible due to the $(m_{\nu}/m_W)^{4}$ suppression, placing it far below the current experimental limits. This makes the conversion an excellent venue for testing new physics as observable rates typically arise in the models, e.g.~\cite{Weinberg:1959zz,Shanker:1979ap,Kitano:2002mt,Cirigliano:2009bz,Deshpande:2011uv,Deppisch:2012vj,deGouvea:2013zba,Crivellin:2014cta,Calibbi:2017uvl,Fornal:2018dqn,Bauer:2021mvw}. 

The current best limits and future sensitivities on branching ratios of the cLFV processes are summarized in Tab.~\ref {exp}. 
\begin{table}[b]
\centering
\begin{tabular}{cc|c}
\hline\hline
Decay modes & Limits on Br. & Exp. \\ 
\hline
$\mu+\text{Ti}\to e+\text{Ti}$ 
& $<6.1\times 10^{-13}$ & SINDRUM II~\cite{PSI1998anu} (1998)\\
$\mu+\text{Au}\to e+\text{Au}$  
& $<7\times 10^{-13}$ & SINDRUM II~\cite{SINDRUMII:2006dvw} (2006)\\
$\mu^+\to e^{+}\gamma$ 
& $<1.5\times 10^{-13}$ & MEG-II~\cite{MEGII:2025gzr} (2025)\\
$\mu^+ \to e^+ e^+ e^-$ &
$<1.0\times 10^{-12}$ & SINDRUM~\cite{SINDRUM:1987nra} (1987)\\
\hline\hline
$\mu+\text{Al}\to e+\text{Al}$ 
& $\sim 3\times 10^{-15} $ &COMET (Phase I)~\cite{COMET:2018auw}\\
$\mu+\text{Al}\to e+\text{Al}$ 
& $\sim 3\times 10^{-17} $ &COMET (Phase II)\\
$\mu+\text{Al}\to e+\text{Al}$ 
& $\sim 6.2\times 10^{-16} $& Mu2e (Run I)~\cite{Miscetti:2025uxk}\\
$\mu+\text{Al}\to e+\text{Al}$ 
& $\sim 3\times 10^{-17} $& Mu2e (Run II)~\cite{Mu2e-II:2022blh}\\
$\mu^+\to e^{+}\gamma$ 
& $\sim6.0\times 10^{-14}$ & MEG-II~\cite{MEGII:2025gzr}\\
$\mu^+ \to e^+ e^+ e^-$ 
& $\sim2.0\times 10^{-15}$ & Mu3e (Phase I)~\cite{Mu3e:2020gyw}\\
$\mu^+ \to e^+ e^+ e^-$ 
& $\sim 10^{-16}$ & Mu3e (Phase II) \\
\hline\hline
\end{tabular}
\caption{The current best limits and the future sensitivities on branching ratios of the charged lepton flavor violation experiments with muons. A good mini-review can be seen in~\cite{COMET:2025sdw}.}
\label{exp}
\end{table}
Searches for $\mu$-$e$ conversion and $\mu\to3e$ decay have not yielded updated experimental limits in twenty years. With the development of experimental technologies, the upcoming experiments, COMET Phase I~\cite{COMET:2018auw} and Mu2e Run I~\cite{Miscetti:2025uxk}, will improve the sensitivity by 2-3 orders of magnitude in the near future, drawing greater attention to muon cLFV and enabling a more comprehensive examination of new physics models, such as the $R$-parity violating Supersymmetric model. 
 
In the Minimal Supersymmetric Standard Model (MSSM), the superpotential related to Yukawa couplings is given by
\begin{align}
	W_{\rm{MSSM}}=y_{u}U^cQH_{u}-y_{d}D^cQH_{d}-y_{e}E^cLH_{d}+\mu H_{u}H_{d}~.
\end{align}
The superfields $Q$, $U$, $D$, $L$, and $E$ contain the Standard Model (SM) fermions, while $H_{u,d}$ are related to the SM Higgs fields. Both $H_u$ and $H_d$ are necessary to avoid gauge anomalies and to generate masses for both up-type and down-type fermions after electroweak symmetry breaking. The above superpotential respects the $R$-parity symmetry~\cite{Weinberg:1981wj,Sakai:1981pk,Dimopoulos:1981dw} defined by
\begin{align}
P_{R}=(-1)^{3B+L+2S}~,
\end{align}
where $S$ is the spin of the particle, $B$ and $L$ are the baryon and lepton number, respectively. Under $R$-parity, all the SM particles and the Higgs bosons have even $R$-parity $(P_{R}=+1)$, while squarks, sleptons, gauginos, and higgsinos have odd $R$-parity $(P_{R}=-1)$. Consequently, the MSSM superpotential $W_{\text{MSSM}} $ is $R$-parity invariant, thus avoiding tree-level cLFV, similar to the Standard Model.

The $R$-parity is introduced to prevent baryon and lepton number violation, thus avoiding rapid proton decay. However, gravitational effects may break this global symmetry, while gauge symmetries are likely more robust. Without violating the SM gauge symmetries, there are other terms that one can write down as
\begin{align}
W\supset\dfrac{1}{2}\lambda_{ijk}L_{i}L_{j}E_{k}^{c}+\lambda_{ijk}^{\prime}L_{i}Q_{j}D_{k}^{c}+\dfrac{1}{2}\lambda_{ijk}^{\prime\prime}U_{i}^{c}D_{j}^{c}D_{k}^{c}+\kappa_{i}L_{i}H_{u}~,
\label{WR}\end{align}
where $i,j,k$ are generation indices. These terms violate R-parity. The Supersymmetric model with $R$-parity violating interactions has rich phenomenology and has been well studied in the literature~\cite{Mohapatra:1993bk,Barbier:2004ez,Chemtob:2004xr,Choudhury:2024ggy,Altmannshofer:2025jkk}. These $R$-parity violating terms can generate neutrino masses~\cite{Hall:1983id,Babu:1989px,Hirsch:2000ef}, and satisfy one of the Sakharov conditions for explaining the universe's matter-antimatter asymmetry. Simultaneously, the terms are highly constrained by low-energy processes, e.g., meson mixing, $K$ and $B$ meson decays, and neutrinoless double beta decay~\cite{Cirigliano:2004tc,Ramsey-Musolf:2006evg,Chen:2007dg,Deshpande:2016yrv,Domingo:2018qfg}. 

The cLFV processes with the bilinear and trilinear $R$-parity violating interactions in Supersymmetric model have been studied extensively in previous literature~\cite{Barbieri:1990qj,Hinchliffe:1992ad,Chaichian:1996wr,Choudhury:1996ia,Kim:1997rr,deGouvea:2000cf,Choi:2000bm,Cheung:2001sb,Carvalho:2002bq,Gemintern:2003gd,Chen:2008gd,Bose:2010eb,Dreiner:2012mx}. In this work, we focus on the effects of trilinear $R$-parity violating interactions on $\mu$-$e$ conversion. We aim for a comprehensive calculation by deriving effective operators and coefficients, performing renormalization group (RG) evolutions down to low energy, and comparing results with experimental data from $\mu$-$e$ conversion, $\mu \to e \gamma$, and $\mu \to eee$ to constrain relevant parameters\footnote{The bilinear term may also have effects on $\mu$-$e$ conversion, we set the $\kappa$ parameter to be zero so that to focus on the trilinear terms}. Our main goal is to comprehensively update the upper limits by those muon cLFV experimental results and sensitivities on the 15 trilinear parameters $\lambda^{\prime}$ in terms of $\lambda^{\prime}_{ijk}\lambda_{i^{\prime}j^{\prime}k^{\prime}}^{\prime*}$ and 6 parameters in terms of $\lambda_{ijk}\lambda_{i^{\prime}j^{\prime}k^{\prime}}^{*}$ , and determine the significance of RG effects.

\section{Operators for \texorpdfstring{$\bm{\mu}$-$\bm{e}$}{} conversion due to \texorpdfstring{$\bm{R}$}{}-parity violation}

Among the three trilinear $R$-parity violating terms, the $\lambda_{ijk}$ and $\lambda'_{ijk}$ terms appear first at the one-loop level and at the tree level, respectively, while the $\lambda^{\prime\prime}_{ijk}$ terms only contribute at higher orders.  We will concentrate on the effects from $\lambda_{ijk}$ and $\lambda'_{ijk}$ terms. As collider searches have constrained the masses of new particles to be larger than $\mathcal{O}$(TeV), it is necessary for us to follow the standard procedure of matching and running within the effective field theory (EFT) framework. 

\subsection{Matching the operators to SMEFT}\label{MC-SM}
One can match the new physics model to the Standard Model effective field theory (SMEFT) at the high new physics scale $\Lambda_{\text{NP}}$ with operators of different dimensions 
\begin{align}
\mathcal{L}_{\text{SMEFT}}=\mathcal{L}_{\text{SM,(4)}}+\sum\limits_{i}\mathcal{C}_{i,(5)}\mathcal{O}_{i,(5)}+\sum\limits_{i}\mathcal{C}_{i,(6)}\mathcal{O}_{i,(6)}+...~,
\end{align}
where the $\mathcal{C}_{i}$ are the corresponding dimensional coefficients determined by integrating out the heavy fields. The dimension-six operators relevant to the muon charged flavor violation processes~\cite{Grzadkowski:2010es,Crivellin:2013hpa,Delzanno:2024ooj} are collected in Tab.~\ref{SMEFT}, where $a,b=1,2$ label the components of the weak isospin doublets and $\tau^{I}~( I=1,2,3 )$ is the Pauli matrices. 
\begin{table}[t]
\centering
\resizebox{\textwidth}{!}{
\renewcommand{\arraystretch}{1.2}
\begin{tabular}{||c|c|c|c|c|c||}
\hline\hline
\multicolumn{2}{||c|}{$\ell \ell D \varphi^{2}$/$\ell \ell  \varphi^{3}$} &
\multicolumn{2}{c|}{$\ell \ell \varphi X$}&
\multicolumn{2}{|c||}{$\ell \ell \ell \ell$} \\
\hline
$\mathcal{O}_{\varphi \ell}^{(1)}$ & $(\varphi^\dagger i\!\overleftrightarrow{D}_\mu \varphi)(\overline{L_{Lp}}\gamma^\mu L_{Lr})$
&  $\mathcal{O}_{eW}$ & $ (\overline{L_{Lp}}\sigma^{\mu\nu} E_{Rr})\tau^I\varphi W^I_{\mu\nu}$
&$\mathcal{O}_{\ell\ell}$ & $(\overline{L_{Lp}}\gamma_\mu L_{Lr})(\overline{L_{Ls}}\gamma^\mu L_{Lt})$ \\
$\mathcal{O}_{\varphi \ell}^{(3)}$ & $(\varphi^\dagger i\!\overleftrightarrow{D}_\mu^I \varphi)(\overline{L_{Lp}} \tau^I\gamma^\mu L_{Lr})$
& $ \mathcal{O}_{eB}$ & $(\overline{\ell}_p\sigma^{\mu\nu} E_{Rr})\varphi B_{\mu\nu}$
&$\mathcal{O}_{ee}$ & $(\overline{E_{Rp}}\gamma_\mu E_{Rr})(\overline{E_{Rs}}\gamma^\mu E_{Rt})$ 
\\ 
$\mathcal{O}_{\varphi e}$ & $(\varphi^\dagger i\!\overleftrightarrow{D}_\mu \varphi)(\overline{E_{Rp}}\gamma^\mu E_{Rr})$ 
&&
& $\mathcal{O}_{\ell e}$ & $(\overline{L_{Lp}}\gamma_\mu L_{Lr})(\overline{E_{Rs}}\gamma^\mu E_{Rt})$\\
$ \mathcal{O}_{e\varphi}$ & $(\varphi^\dagger\varphi)(\overline{L_{Lp}} E_{Rr}\varphi)$ &&&& \\ 
\hline\hline
\multicolumn{6}{|c|}{$\ell \ell q q$}\\
\hline
$\mathcal{O}_{\ell q}^{(1)}$ & $(\overline{L_{Lp}}\gamma_\mu L_{Lr})(\overline{Q_{Ls}}\gamma^\mu Q_{Lt})$ 
&$\mathcal{O}_{\ell d}$ & $(\overline{L_{Lp}}\gamma_\mu L_{Lr})(\overline{D_{Rs}}\gamma^\mu D_{Rt})$
&$\mathcal{O}_{\ell u}$ & $(\overline{L_{Lp}}\gamma_\mu L_{Lr})(\overline{U_{Rs}}\gamma^\mu U_{Rt})$  \\
$\mathcal{O}_{\ell q}^{(3)}$ & $(\overline{L_{Lp}}\gamma_\mu \tau^I L_{Lr})(\overline{Q_{Ls}}\gamma^\mu \tau^I Q_{Lt})$ 
&$\mathcal{O}_{e d}$ & $(\overline{E_{Rp}}\gamma_\mu E_{Rr})(\overline{D_{Rs}}\gamma^\mu D_{Rt})$
& $\mathcal{O}_{e u}$ & $(\overline{E_{Rp}}\gamma_\mu E_{Rr})(\overline{U_{Rs}}\gamma^\mu U_{Rt})$\\
$\mathcal{O}_{\ell e q u}^{(1)}$ &  $(\overline{L_{Lp}^{a}}E_{Rr})\epsilon_{ab}(\overline{Q_{Ls}^{b}}U_{Rt})$
&$\mathcal{O}_{\ell e d q}$ &  $(\overline{L_{Lp}^{a}} E_{Rr})(\overline{D_{Rs}} Q_{Lt}^{a})$
&$\mathcal{O}_{qe}$ & $(\overline{Q_{Lp}}\gamma_\mu Q_{Lr})(\overline{E_{Rs}}\gamma^\mu E_{Rt})$  \\ 
$\mathcal{O}_{\ell e q u}^{(3)}$ & $(\overline{L_{Lp}^{a}}\sigma_{\mu\nu} E_{Rr})\epsilon_{ab}(\overline{Q_{Ls}^{b}}\sigma^{\mu\nu} U_{Rt})$&&&& \\
\hline\hline
\end{tabular}}
\caption{The SMEFT operators that are relevant to the muon charged flavor violation processes.}\label{SMEFT}
\end{table}
We keep the four-fermion operators in the table labeled ``$\ell\ell\ell\ell$'' which can directly contribute to $\mu\to 3e$ for completeness.

For the $R$-parity violating SUSY model we are considering, the $\lambda_{ijk}$ and $\lambda'_{ijk}$ terms in the superpotential as shown in Eq. (\ref{WR}) will induce the following interactions
\begin{align}
	\mathcal{L}_{\lambda}
	=-\dfrac{1}{2}\lambda_{ijk}
	\bigg(&\overline{L_{Li}^{c}}\epsilon L_{Lj}\tilde{E}_{Rk}^{*}
	+\overline{L_{Li}^{c}}\epsilon \tilde{L}_{Lj}E_{Rk}^{c}
	+\tilde{L}_{Li}^{*T}\epsilon \overline{L_{Lj}^{c}}^{T}E_{Rk}^{c}\bigg)+\text{h.c.}\notag\\
	=-\lambda_{ijk} 
	\bigg(&\overline{\nu_{Li}^{c}}E_{Lj}\tilde{E}_{Rk}^{*}
	+\overline{\nu_{Li}^{c}}\tilde{E}_{Lj}E_{Rk}^{c}
	+\tilde{\nu}_{Li}\overline{E_{Lj}^{c}}E_{Rk}^{c}\bigg)+\text{h.c.}~,\label{lambda-Lag}\\
	\mathcal{L}_{\lambda^{\prime}}
	=-\lambda_{ijk}^{\prime}
	\bigg(&\overline{L_{Li}^{c}}\epsilon Q_{Lj} \tilde{D}_{Rk}^{*}
	+\overline{L_{Li}^{c}}\epsilon \tilde{Q}_{Lj} D_{Rk}^{c}
	+\tilde{L}_{Li}^{T}\epsilon \overline{Q_{Lj}^{c}} D_{Rk}^{c}\bigg)+\text{h.c.}\notag\\
	=-\lambda_{ijk}^{\prime}
	\bigg(&\overline{\nu_{Li}^{c}} D_{Lj} \tilde{D}_{Rk}^{*}
	+\overline{\nu_{Li}^{c}}\tilde{D}_{Lj} D_{Rk}^{c}
	+\tilde{\nu}_{Li}\overline{D_{Lj}^{c}} D_{Rk}^{c}\notag\\
	-&\overline{E_{Li}^{c}} U_{Lj} \tilde{D}_{Rk}^{*}
	-\overline{E_{Li}^{c}}\tilde{U}_{Lj} D_{Rk}^{c}
	-\tilde{E}_{Li}\overline{U_{Lj}^{c}} D_{Rk}^{c}
	\bigg)+\text{h.c.}~.\label{lambdap-Lag}
\end{align}
where the $\tilde{f}$ are the s-fermion fields for the corresponding SM fermion $f$ ones. From here on, we will take the notations $Q,U,D,L,E$ to be the SM fermions
\begin{align}
\begin{gathered}
Q_{Li}=\begin{pmatrix}U_{Li}\\D_{Li}\end{pmatrix}
\equiv\begin{pmatrix}u_L,~c_L,~t_L\\d_L,~s_L,~b_L\end{pmatrix}~,
\quad L_{Li}=\begin{pmatrix}\nu_{Li}\\e_{Li}\end{pmatrix}
\equiv\begin{pmatrix}\nu_{eL},~\nu_{\mu L},~\nu_{\tau L} \\e_L,~\mu_L,~\tau_L\end{pmatrix}~,\\
U_{Ri}\equiv (u_R,~c_R,~t_R)~,\quad
D_{Ri}\equiv (d_R,~s_R,~b_R)~,\quad
E_{Ri}\equiv (e_R,~\mu_R,~\tau_R)~,
\end{gathered}
\end{align}
and the charge conjugate in Eq.~(\ref{lambda-Lag}) and (\ref{lambdap-Lag}) are defined as $\nu_{L}^{c}=(\nu_L)^c, E_{R}^c=(E_{R})^c$ and etc..
Note that the coefficient $\lambda_{ijk}$ satisfies the anti-symmetry relation $\lambda_{ijk}=-\lambda_{jik}$. Then one can integrate out the heavy s-fermion fields and determine the coefficients.
For the $\lambda'_{ijk}$ term, the effective Lagrangian at the tree level contains
\begin{align}
\mathcal{L}_{\text{eff}}
&\supset\dfrac{\lambda_{ijk}^{\prime}\lambda_{mnk}^{\prime*}}{2m^{2}_{\tilde{D}_{Rk}}}
[\overline{D_{Ln}}\gamma_{\mu}D_{Lj}][\overline{\nu_{Lm}}\gamma^{\mu}\nu_{Li}]
+\dfrac{\lambda_{ijk}^{\prime}\lambda_{mnk}^{\prime*}}{2m^{2}_{\tilde{D}_{Rk}}}[\overline{U_{Ln}}\gamma_{\mu}U_{Lj}][\overline{E_{Lm}}\gamma^{\mu}E_{Li}]\notag\\
&-\dfrac{\lambda_{ijk}^{\prime}\lambda_{mnk}^{\prime*}}{2m^{2}_{\tilde{D}_{Rk}}}[\overline{D_{Ln}}\gamma_{\mu}U_{Lj}][\overline{\nu_{Lm}}\gamma^{\mu}E_{Li}]
-\dfrac{\lambda_{ijk}^{\prime}\lambda_{mjn}^{\prime*}}{2m^{2}_{\tilde{D}_{Lj}}}[\overline{D_{Rk}}\gamma_{\mu}D_{Rn}][\overline{\nu_{Lm}}\gamma^{\mu}\nu_{Li}]\notag\\
&-\dfrac{\lambda_{ijk}^{\prime}\lambda_{mjn}^{\prime*}}{2m_{\tilde{U}_{Rk}}^{2}}[\overline{D_{Rk}}\gamma_{\mu}D_{Rn}][\overline{E_{Lm}}\gamma^{\mu}E_{Li}]+\text{h.c.}~.
\end{align}
These operators can directly contribute to the $\mu$-$e$ conversion process by exchanging an s-fermion at the tree level. After doing the linear combinations, these operators can be converted to those SMEFT operators with the ``$\ell\ell qq$'' tag. Then one can give the matching conditions at the new physics scale $\Lambda_{\text{NP}}$ as
\begin{align}
[\mathcal{C}_{\ell q}^{(1)}]_{prst}(\Lambda_{\text{NP}})=\dfrac{\lambda_{rtk}^{\prime}\lambda_{psk}^{\prime *}}{4m_{\tilde{D}_{Rk}}^{2}}~,~
[\mathcal{C}_{\ell q}^{(3)}]_{prst}(\Lambda_{\text{NP}})=-\dfrac{\lambda_{rtk}^{\prime}\lambda_{psk}^{\prime *}}{4m_{\tilde{D}_{Rk}}^{2}}~,~
[\mathcal{C}_{\ell d}]_{prst}(\Lambda_{\text{NP}})=-\dfrac{\lambda_{rjs}^{\prime}\lambda_{pjt}^{\prime *}}{2m_{\tilde{U}_{Lj}}^{2}}~.
\label{lpSMEFT}
\end{align}

For the $\lambda_{ijk}$ term, it is not possible to contribute the $\mu$-$e$ conversion process at the tree level. However, one can generate a radiative contribution with the photon $\gamma$ and the $Z$ boson attaching to quarks and induce the conversion process. The corresponding SMEFT operators that contain $\gamma$ or $Z$ explicitly are $\mathcal{O}_{eW,eB}$ and $\mathcal{O}_{\varphi \ell}^{(1,3)},\mathcal{O}_{\varphi e}$ where the first two operators can not only contribute to $\mu$-$e$ conversion but also to the radiative decay $\mu\to e\gamma$. 
The vector type four-fermion operators as shown in Tab.~\ref{SMEFT} can also be obtained from the one-loop diagrams and be transformed by the equation of motions (EOMs)~\cite{Grzadkowski:2010es}. The matching results can be easily derived by computing the radiative diagrams and being convinced by the \texttt{Mathematica} package \texttt{Matchete}~\cite{Fuentes-Martin:2022jrf}. Here we show the $\mathcal{O}_{eW,eB}$ Wilson coefficients that relate with $\lambda_{32k}\lambda_{31k}^{*}$ as an example
\begin{align}
\mathcal{C}_{eW}^{12}(\Lambda_{\text{NP}})&=\dfrac{g_{2}Y_{e}^{s2}}{128\pi^{2}m_{\tilde{\ell}}^{2}}
\lambda_{3sk}\lambda_{31k}^{*}(F_{2}^{(1)}-F_{5}^{(2)})~,\\
\mathcal{C}_{eB}^{12}(\Lambda_{\text{NP}})&=-\dfrac{g_{1}Y_{e}^{s2}}{128\pi^{2}m_{\tilde{\ell}}^{2}}
\lambda_{3sk}\lambda_{31k}^{*}(F_{2}^{(1)}-F_{5}^{(2)})~,\label{lSMEFT-1}\\
\mathcal{C}_{eW}^{21}(\Lambda_{\text{NP}})&=\dfrac{g_{2}Y_{e}^{s1}}{128\pi^{2}m_{\tilde{\ell}}^{2}}
\lambda_{32k}\lambda_{3sk}^{*}(F_{2}^{(1)}-F_{5}^{(2)})~,\\
\mathcal{C}_{eB}^{21}(\Lambda_{\text{NP}})&=-\dfrac{g_{1}Y_{e}^{s1}}{128\pi^{2}m_{\tilde{\ell}}^{2}}
\lambda_{32k}\lambda_{3sk}^{*}(F_{2}^{(1)}-F_{5}^{(2)})~.\label{lSMEFT-2}
\end{align}
with $g_{1,2}$ to be the couplings of $U(1)_{Y}$ and $SU(2)_{L}$, respectively. The loop functions are
\begin{align}
F_{2}=\dfrac{2t^{2}+5t-1}{6(t-1)^{3}}-\dfrac{t^{2}\ln t}{(t-1)^{4}}~,\quad
F_{5}=\dfrac{t^{2}-5t-2}{6(t-1)^{3}}+\dfrac{t\ln t}{(t-1)^{4}}~,
\end{align}
with $t=m_{\nu_{3}}^{2}/m_{\tilde{\ell}}^{2}$ in $F_{2,5}^{(1)}$ and $t=m_{E_{Rk}}^{2}/m_{\tilde{\ell}}^{2}$ in $F_{2,5}^{(2)}$. Here we take all the slepton masses $m_{\tilde{E}_{Rk}}=m_{\tilde{E}_{Lj}}=m_{\tilde{\nu}_{Lj}}\equiv m_{\tilde{\ell}}\sim \mathcal{O}$(TeV), then at the limit $t\to 0$, one can derive that $F_{5}\simeq 1/3$ and $F_{2}=1/6$. 


After the electroweak symmetry breaking, the gauge bosons $W,Z$, the Higgs boson $\varphi$ in the SMEFT should be integrated out. We then deal with the processes in the low-energy effective field theory (LEFT). In the following discussion, the matching conditions between SMEFT and LEFT at the electroweak scale $\Lambda_{\text{EW}}$ are shown.

\subsection{Matching the operators to LEFT}
For the discussion at the low-energy scale, we follow the standard basis in~\cite{Jenkins:2017jig}, and the related LEFT operators that contribute to the $\mu$-$e$ conversion process are
\begin{align}
\mathcal{L}_{\text{eff,d5}}^{\text{LEFT}}
&=[\mathcal{C}_{e\gamma}]_{pr}(\overline{E_{Lp}}\sigma^{\mu\nu}E_{Rr})F_{\mu\nu}~,\label{dim5}\\
\mathcal{L}_{\text{eff,d6}}^{\text{LEFT}}
&=[\mathcal{C}_{eu}^{V,LL}]_{prst}
(\overline{E_{Lp}}\gamma^{\mu}E_{Lr})(\overline{U_{Ls}}\gamma_{\mu}U_{Lt})
+[\mathcal{C}_{ed}^{V,LL}]_{prst}
(\overline{E_{Lp}}\gamma^{\mu}E_{Lr})(\overline{D_{Ls}}\gamma_{\mu}D_{Lt})\notag\\
&+[\mathcal{C}_{eu}^{V,LR}]_{prst}
(\overline{E_{Lp}}\gamma^{\mu}E_{Lr})(\overline{U_{Rs}}\gamma_{\mu}U_{Rt})
+[\mathcal{C}_{ed}^{V,LR}]_{prst}
(\overline{E_{Lp}}\gamma^{\mu}E_{Lr})(\overline{D_{Rs}}\gamma_{\mu}D_{Rt})\notag\\
&+[\mathcal{C}_{ue}^{V,LR}]_{prst}
(\overline{U_{Lp}}\gamma^{\mu}U_{Lr})(\overline{E_{Rs}}\gamma_{\mu}E_{Rt})
+[\mathcal{C}_{de}^{V,LR}]_{prst}
(\overline{D_{Lp}}\gamma^{\mu}D_{Lr})(\overline{E_{Rs}}\gamma_{\mu}E_{Rt})\notag\\
&+[\mathcal{C}_{eu}^{V,RR}]_{prst}
(\overline{E_{Rp}}\gamma^{\mu}E_{Rr})(\overline{U_{Rs}}\gamma_{\mu}U_{Rt})
+[\mathcal{C}_{ed}^{V,RR}]_{prst}
(\overline{E_{Rp}}\gamma^{\mu}E_{Rr})(\overline{D_{Rs}}\gamma_{\mu}D_{Rt})\notag\\
&+[\mathcal{C}_{eu}^{S,RR}]_{prst}
(\overline{E_{Lp}}E_{Rr})(\overline{U_{Ls}}U_{Rt})
+[\mathcal{C}_{ed}^{S,RR}]_{prst}
(\overline{E_{Lp}}E_{Rr})(\overline{D_{Ls}}D_{Rt})\notag\\
&+[\mathcal{C}_{eu}^{S,RL}]_{prst}
(\overline{E_{Lp}}E_{Rr})(\overline{U_{Rs}}U_{Lt})
+[\mathcal{C}_{ed}^{S,RL}]_{prst}
(\overline{E_{Lp}}E_{Rr})(\overline{D_{Rs}}D_{Lt})\notag\\
&+[\mathcal{C}_{eu}^{T,RR}]_{prst}
(\overline{E_{Lp}}\sigma^{\mu\nu}E_{Rr})(\overline{U_{Ls}}\sigma_{\mu\nu}U_{Rt})\notag\\
&+[\mathcal{C}_{ed}^{T,RR}]_{prst}
(\overline{E_{Lp}}\sigma^{\mu\nu}E_{Rr})(\overline{D_{Ls}}\sigma_{\mu\nu}D_{Rt})~.\label{dim62}
\end{align}
These Lagrangians and operators are in the flavor basis, whereas numerical analysis should be performed in the mass eigenstates. Here we choose the basis where down-type quarks are already diagonalized, and the  $U_{L\alpha}$ should be replaced by 
$(V^\dagger_{\text{CKM}})_{\alpha i} U_{Li}$ in the above effective Lagrangian, where the $V_{\text{CKM}}$ is the Cabibbo-Kobayashi-Maskawa (CKM) matrix with the standard parameterization, and the values are chosen from the \texttt{python} package \texttt{wilson}~\cite{Aebischer:2018bkb}
\begin{align}
V_{\text{CKM}}=
\begin{pmatrix}
0.9745 & 0.2243 & (1.236-3.402 i)\times 10^{-3}\\
-0.2242-1.399\times 10^{-4} i & 0.9736-3.221\times 10^{-4} i & 0.04220\\
(8.262-3.313i)\times 10^{-3} & -0.04140-7.625\times 10^{-4} i & 0.9991
\end{pmatrix}~.
\end{align}

The dimension-five operators in Eq.~(\ref{dim5}) can not only contribute to the $\mu$-$e$ conversion, but also contribute to muon radiative decay. The contributions from the $Z$ boson are not explicit as in e.g.~\cite{Hisano:1995cp,Arganda:2005ji,Dreiner:2012mx} have discussed, because it has been integrated out when the theory is at the 2 GeV scale, which is far below the scale of the electroweak symmetry breaking. The matching conditions from SMEFT to LEFT at the electroweak scale $\Lambda_{\text{EW}}$ have been completely derived in~\cite{Jenkins:2017jig}, and here we show the conditions corresponding to $\mathcal{C}_{\ell q}^{(1,3)}, \mathcal{C}_{\ell d}$ as examples
\begin{gather}
[\mathcal{C}_{eu}^{V,LL}]_{prst}(\Lambda_{\text{EW}})=[\mathcal{C}_{\ell q}^{(1)}-\mathcal{C}_{\ell q}^{(3)}]_{prst}(\Lambda_{\text{EW}})+\mathcal{C}_{Z}(g_{Z}^{2}/m_{Z}^{2})~,\\
[\mathcal{C}_{ed}^{V,LL}]_{prst}(\Lambda_{\text{EW}})= [\mathcal{C}_{\ell q}^{(1)}+\mathcal{C}_{\ell q}^{(3)}]_{prst}(\Lambda_{\text{EW}})+\mathcal{C}_{Z}(g_{Z}^{2}/m_{Z}^{2})~,\\
[\mathcal{C}_{ed}^{V,LR}]_{prst}(\Lambda_{\text{EW}})=[\mathcal{C}_{\ell d}]_{prst}(\Lambda_{\text{EW}})+\mathcal{C}_{Z}(g_{Z}^{2}/m_{Z}^{2})~,
\end{gather}
where the $\mathcal{C}_{Z}$ represents the $Z$ boson contributions. For the operators $\mathcal{O}_{eW,eB}$, the matching conditions are 
\begin{align}
\mathcal{C}_{e\gamma,pr}(\Lambda_{\text{EW}})=\dfrac{v}{\sqrt{2}}[-\mathcal{C}_{eW}^{pr}(\Lambda_{\text{EW}})\bar{s}+\mathcal{C}_{eB}^{pr}(\Lambda_{\text{EW}})\bar{c}]~,
\end{align}
where $v=246$ GeV is the vacuum expected value and $\bar{s}=\sin{\theta_{W}},\bar{c}=\cos{\theta_{W}}$ with $\theta_{W}$ to be the weak mixing angle. 

\subsection{RG running effects}
Following the standard steps of the analysis, one should also perform the renormalization group (RG) evolutions of the SMEFT Wilson coefficients from $\Lambda_{\text{NP}}$ to $\Lambda_{\text{EW}}$~\cite{Jenkins:2013zja,Jenkins:2013wua,Alonso:2013hga}
\begin{align}
\frac{d \mathcal{C}_{i}}{d \ln \mu}=\dfrac{1}{16 \pi^2} \sum_{j} \gamma_{j i} \mathcal{C}_{i}~,
\end{align}
and RG effects of the LEFT Wilson coefficients from $\Lambda_{\text{EW}}$ to $\Lambda_{\text{low}}=$ 2 GeV~\cite{Jenkins:2017dyc} which is relevant for low-energy precision tests
 \begin{align}
 \frac{d \mathcal{C}_{i}}{d \ln \mu}=\dfrac{g_{s}^{2}}{16 \pi^2} \sum_{j} \gamma^{s}_{j i} \mathcal{C}_{i}+\dfrac{e^{2}}{16 \pi^2} \sum_{j} \gamma^{e}_{j i} \mathcal{C}_{i}~,
 \end{align}
with $\gamma,\gamma^{s,e}$ to be the anomalous dimension matrices. These RG running effects can be automatically solved by using the \texttt{python} package \texttt{wilson}~\cite{Aebischer:2018bkb}. In the numerical analysis, the RG effects of relevant operators are solved from above TeV SMEFT in Warsaw-up basis to 2 GeV LEFT in WET basis, which are shown in the Tab.~\ref{RGE}.
\begin{table}[t]
\resizebox{\textwidth}{!}{
\renewcommand{\arraystretch}{1.2}
\begin{tabular}{|c|c|c|c|c|c|c|c|c|c|}
\hline\hline
\multirow{3}{*}{Para.} 
& \multicolumn{2}{|c|}{$\mathcal{C}_{eu,1211}^{V,LL}$} 
& \multicolumn{2}{|c|}{$\mathcal{C}_{ed,1211}^{V,LL}$}
& \multicolumn{2}{|c|}{$\mathcal{C}_{eu,1211}^{V,LR}$}
& \multicolumn{2}{|c|}{$\mathcal{C}_{ed,1211}^{V,LR}$} \\
\cline{2-9}
& w/  & w/o 
& w/  & w/o 
& w/  & w/o 
& w/  & w/o \\
& RG effects & RG effects
& RG effects & RG effects
& RG effects & RG effects
& RG effects & RG effects\\
\hline
$\lambda_{211}^{\prime}\lambda_{111}^{\prime *}$ 
& $0.5130$               & $0.4748$
& $2.768\times10^{-3}$  & $0$
& $-5.735\times10^{-3}$ & $0$
& $-0.5045$              & $-0.5$\\
$\lambda_{212}^{\prime}\lambda_{112}^{\prime *}$ 
& $0.5130$               & $0.4748$
& $2.768\times10^{-3}$  & $0$
& $-5.735\times10^{-3}$ & $0$
& $2.908\times10^{-3}$  & $0$\\
$\lambda_{213}^{\prime}\lambda_{113}^{\prime *}$ 
& $0.5130$               & $0.4748$
& $2.758\times10^{-3}$  & $0$
& $-5.735\times10^{-3}$ & $0$
& $2.908\times10^{-3}$  & $0$\\
$\lambda_{221}^{\prime}\lambda_{121}^{\prime *}$ 
& $2.147\times10^{-2}$  & $2.515\times10^{-2}$
& $2.828\times10^{-3}$  & $0$
& $-5.708\times10^{-3}$ & $0$
& $-0.5045$              & $-0.5$\\
$\lambda_{231}^{\prime}\lambda_{131}^{\prime *}$ 
& $-3.352\times10^{-2}$             &  $0$
& $3.743\times10^{-2}$              &  $0$
& $1.016\times10^{-2}$  &  $0$
& $-0.5128$              &  $-0.5$\\
\hline
$\lambda_{222}^{\prime}\lambda_{122}^{\prime *}$ 
& $2.147\times10^{-2}$   & $2.515\times10^{-2}$
& $2.828\times10^{-3}$   & $0$
& $-5.708\times10^{-3}$  & $0$
& $2.895\times10^{-3}$   & $0$\\
$\lambda_{223}^{\prime}\lambda_{123}^{\prime *}$ 
& $2.148\times10^{-2}$   & $2.515\times10^{-2}$
& $2.818\times10^{-3}$   & $0$
& $-5.710\times10^{-3}$  & $0$
& $2.895\times10^{-3}$   & $0$\\
$\lambda_{22k}^{\prime}\lambda_{11k}^{\prime *}$
& $0.1195$ & $0.1093$
& 0 & 0
& 0 & 0
& 0 & 0\\
$\lambda_{21k}^{\prime}\lambda_{12k}^{\prime *}$ 
& $0.1195$ & $0.1093$
& $0$ & $0$
& $0$ & $0$
& $0$ & $0$\\
$\lambda_{23k}^{\prime}\lambda_{11k}^{\prime *}$ 
& $3.448\times10^{-4}$ & $5.215\times10^{-4}$
& $2.988\times10^{-4}$ & $0$
& $1.370\times10^{-4}$ & $0$
& $7.028\times10^{-5}$ & $0$\\
$\lambda_{21k}^{\prime}\lambda_{13k}^{\prime *}$ 
& $3.448\times10^{-4}$ & $5.215\times10^{-4}$
& $2.988\times10^{-4}$ & $0$
& $1.370\times10^{-4}$ & $0$
& $7.028\times10^{-5}$ & $0$\\
$\lambda_{23k}^{\prime}\lambda_{12k}^{\prime *}$ 
& $1.301\times10^{-3}$ & $1.201\times10^{-4}$
& $-1.468\times10^{-3}$  & 0
& $-6.730\times10^{-4}$  & 0
& $3.450\times10^{-4}$ & 0\\
$\lambda_{22k}^{\prime}\lambda_{13k}^{\prime *}$ 
& $1.301\times10^{-3}$ & $1.201\times10^{-4}$
& $-1.468\times10^{-3}$  & 0
& $-6.730\times10^{-4}$  & 0
& $3.450\times10^{-4}$ & 0\\
\hline\hline
\end{tabular}}
\caption{The numerical expressions for the Wilson coefficients of LEFT operators with or without RG running effects that are related to $\mu$-$e$ conversion, which can be tree-level contributed from the $\lambda^{\prime}$ combinations. One can see the text for the details.}
\label{RGE}
\end{table}
The Wilson coefficients of LEFT operators can be numerically derived in the form of the numbers in the table multiplied by the combinations $\lambda^{\prime}\lambda^{\prime*}/m_{\tilde{q}}^{2}$. For example, if only the combination $\lambda_{211}^{\prime}\lambda_{111}^{\prime*}\neq 0$, the Wilson coefficients without RG running effects are 
\begin{align}  
&[\mathcal{C}_{eu}^{V,LL}]_{1211}
=0.4748\times\dfrac{\lambda_{211}^{\prime}\lambda_{111}^{\prime *}}{m_{\tilde{q}}^{2}}~,\quad
[\mathcal{C}_{eu}^{V,LR}]_{1211}=0~,\notag\\
&[\mathcal{C}_{ed}^{V,LL}]_{1211}=0~,\quad
[\mathcal{C}_{ed}^{V,LR}]_{1211}=-0.5\times\dfrac{\lambda_{211}^{\prime}\lambda_{111}^{\prime *}}{m_{\tilde{q}}^{2}}~,
\end{align}
and with RG running effects, the coefficients become
\begin{align}
&[\mathcal{C}_{eu}^{V,LL}]_{1211}
=0.5130\times\dfrac{\lambda_{211}^{\prime}\lambda_{111}^{\prime *}}{m_{\tilde{q}}^{2}}~,~~
[\mathcal{C}_{eu}^{V,LR}]_{1211}=2.768\times10^{-3}\times\dfrac{\lambda_{211}^{\prime}\lambda_{111}^{\prime *}}{m_{\tilde{q}}^{2}}~,\notag\\
&[\mathcal{C}_{ed}^{V,LL}]_{1211}=-5.735\times10^{-3}\times\dfrac{\lambda_{211}^{\prime}\lambda_{111}^{\prime *}}{m_{\tilde{q}}^{2}}~,~~
[\mathcal{C}_{ed}^{V,LR}]_{1211}=-0.5045\times\dfrac{\lambda_{211}^{\prime}\lambda_{111}^{\prime *}}{m_{\tilde{q}}^{2}}~.
\end{align}
From the Tab.~\ref{RGE}, one can find that RG effects significantly affect the last two combinations, increasing the coefficient $\mathcal{C}_{eu,1211}^{V,LL}$ by an order of magnitude, which will strongly affect the final upper limits and sensitivities given by $\mu$-$e$ conversion experiments. More details will be discussed in the numerical results.

With the high-energy new physics model now analyzed in terms of the low-energy effective operator, we will focus on the phenomenological process of $\mu$-$e$ conversion in the next section, presenting the formula at the low-energy scale within different LEFT bases.

\section{Formula for \texorpdfstring{$\bm{\mu}$-$\bm{e}$}{} conversion}

When a muon beam interacts with a target, some muons are captured by the target nuclei. These captured muons can then interact with nucleons, resulting in their direct conversion to electrons $\mu^{-}  + (A,Z) \to e^{-} + (A,Z)$, where $A$ and $Z$ represent the atomic mass number and the proton number, respectively. The branching ratio characterizes this conversion
\begin{align}
\text{Br}(\mu N \to e N) = \frac{\Gamma_{\text{conv.}}[\mu^{-} + (A,Z) \to e^{-} + (A,Z)]}{\Gamma_{\text{capture}}[\mu^{-} + (A,Z) \to \nu_{\mu} + (A,Z-1)]}~,
\end{align}
where the numerator denotes the $\mu$-$e$ conversion rate within nuclei, and $\Gamma_{\text{capture}}$ is the capture rate of muons resulting in their transformation to muon neutrinos. This formalism is crucial for understanding cLFV processes in atomic systems. 

\subsection{The conversion rate}\label{3.1}
Following the basis in Eq. (\ref{dim5},\ref{dim62})  at the low-energy scale, the conversion rate related to spin-independent can be written as
\begin{align}
\Gamma_{\text {conv.}}
&= 
\big|(\mathcal{C}_{eu,1211}^{S,RR}+\mathcal{C}_{eu,1211}^{S,RL})(G_{S}^{p,u}S^{(p)}+G_{S}^{n,u}S^{(n)})
+(\mathcal{C}_{ed,1211}^{S,RR}+\mathcal{C}_{ed,1211}^{S,RL})(G_{S}^{p,d}S^{(p)}+G_{S}^{n,d}S^{(n)})\notag\\
&+(\mathcal{C}_{ed,1222}^{S,RR}+\mathcal{C}_{ed,1222}^{S,RL})(G_{S}^{p,s}S^{(p)}+G_{S}^{n,s}S^{(n)})
+(\mathcal{C}_{eu,1211}^{V,LL}+\mathcal{C}_{eu,1211}^{V,LR})(G_{V}^{p,u}V^{(p)}+G_{V}^{n,u}V^{(n)})\notag\\
&+(\mathcal{C}_{ed,1211}^{V,LL}+\mathcal{C}_{ed,1211}^{V,LR})(G_{V}^{p,d}V^{(p)}+G_{V}^{n,d}V^{(n)})
+(\mathcal{C}_{ed,1222}^{V,LL}+\mathcal{C}_{ed,1222}^{V,LR})(G_{V}^{p,s}V^{(p)}+G_{V}^{n,s}V^{(n)})\notag\\
&+D C_{e\gamma,12}^{*}/(2m_{\mu})\big|^{{2}}
+\big|(\mathcal{C}_{eu,2111}^{S,RR*}+\mathcal{C}_{eu,2111}^{S,RL*})(G_{S}^{p,u}S^{(p)}+G_{S}^{n,u}S^{(n)})\notag\\
&+(\mathcal{C}_{ed,2111}^{S,RR*}+\mathcal{C}_{ed,2111}^{S,RL*})(G_{S}^{p,d}S^{(p)}+G_{S}^{n,d}S^{(n)})
+(\mathcal{C}_{ed,2122}^{S,RR*}+\mathcal{C}_{ed,2122}^{S,RL*})(G_{S}^{p,s}S^{(p)}+G_{S}^{n,s}S^{(n)})\notag\\
&+(\mathcal{C}_{eu,1211}^{V,RR}+\mathcal{C}_{ue,1112}^{V,LR})(G_{V}^{p,u}V^{(p)}+G_{V}^{n,u}V^{(n)})
+(\mathcal{C}_{ed,1211}^{V,RR}+\mathcal{C}_{de,1112}^{V,LR})(G_{V}^{p,d}V^{(p)}+G_{V}^{n,d}V^{(n)})\notag\\
&+(\mathcal{C}_{ed,1222}^{V,RR}+\mathcal{C}_{de,2212}^{V,LR})(G_{V}^{p,s}V^{(p)}+G_{V}^{n,s}V^{(n)})
+D C_{e\gamma,21}/(2m_{\mu}) \big|^{{2}}~,
\end{align}
with the vector charges are $G_{V}$ with values to be 
\begin{align}
G_V^{p,u}=G_V^{n,d}=2~,\quad G_V^{p,d}=G_V^{n,u}=1~,\quad  G_V^{p,s}=G_V^{n,s}=0~,\label{GV}
\end{align}
and the scalar charge $G_{S}$ values be
\begin{align}
N_{f}=2+1~,\quad
&G_{S}^{p,u}=6.6~,\quad G_{S}^{p,d}=5.6~,\quad G_{S}^{p,s}=0.47~,\notag\\
&G_{S}^{n,u}=5.5~,\quad G_{S}^{n,d}=6.6~,\quad G_{S}^{n,s}=0.47~;\label{GS1}\\
N_{f}=2+1+1~,\quad
&G_{S}^{p,u}=9.3~,\quad G_{S}^{p,d}=8.3~,\quad G_{S}^{p,s}=0.43~,\notag\\
&G_{S}^{n,u}=8.2~,\quad G_{S}^{n,d}=9.3~,\quad G_{S}^{n,s}=0.43~.\label{GS2}
\end{align}
 The values of these $G_{S}$ factors derived from the latest lattice average results~\cite{FlavourLatticeAveragingGroupFLAG:2024oxs} with $\sigma_{\pi N}=42.2$~MeV, $\sigma_{s}=44.9$~MeV for $N_{f}=2+1$ and $\sigma_{\pi N}=60.9$~MeV, $\sigma_{s}=41.0$~MeV for $N_{f}=2+1+1$. The central values of the $\overline{\text{MS}}$ quark masses are $m_u=2.2~\text{MeV}$, $m_d=4.7~\text{MeV}$, and $m_s=96~\text{MeV}$ which are taken at $\mu=2$~GeV.
The parameters $D, S^{(p),(n)}$,$V^{(p),(n)}$ are overlap integrals, containing the information about the structure of the nucleus. Once the nuclear distributions are specified, the integrals can be calculated. One should be noted that the integral values have uncertainties due to the nuclear distributions. In the numerical analysis, we take the central values of these integrals as benchmark which are from~\cite{Heeck:2022wer}, and the capture rates $\Gamma_{\text{cap.}}$ with the unit to be $10^{6}~\text{s}^{-1}$ in different isotopes~\cite{Suzuki:1987jf} are concluded in Tab.~\ref{values-isotopes}. 
\begin{table}[t]
\centering
 \renewcommand{\arraystretch}{1.2}
\begin{tabular}{c|ccccc|cc}
\hline
\multirow{2}*{Isotopes} & $D$ & $S^{(p)}$ & $V^{(p)}$ &$S^{(n)}$  & $V^{(n)} $ & $\Gamma_{\text{cap.}} $ 
\\
& $[m_{\mu}^{5/2}]$ & $[m_{\mu}^{5/2}]$ & $[m_{\mu}^{5/2}]$  & $[m_{\mu}^{5/2}]$ & $[m_{\mu}^{5/2}]$ & $[10^{6}~\text{s}^{-1}]$ \\
\hline
$^{27}$Al & 0.0359 & 0.0159 &  0.0165 & 0.0172 & 0.0178 & 0.7054 \\
\hline
$^{48}$Ti & 0.0859 & 0.0379 & 0.0407 & 0.0448 & 0.0481 & 2.590 \\
\hline
$^{197}$Au & 0.166 & 0.0523 & 0.0866 & 0.0781 & 0.129& 13.07 \\
\hline\hline
\end{tabular}
\caption{The values of overlap integrals with the unit be $m_{\mu}^{5/2}$ and the capture rate of muons $\Gamma_{\text{cap.}}$ with the unit to be $10^{6}~\text{s}^{-1}$ within the isotopes $^{27}$Al, $^{48}$Ti and $^{197}$Au. The integral uncertainties are about $2\%\sim5\%$ for low $Z$ isotopes $^{27}$Al and $^{48}$Ti, and about $8\%\sim10\%$ for the high $Z$ isotope $^{197}$Au~\cite{Heeck:2022wer}.}
\label{values-isotopes}
\end{table}

\subsection{Transformation between different bases}
Here, we give the transformations between different LEFT bases. As in the previous work, the effective Lagrangian relating to $\mu$-$e$ conversion is usually written as \cite{Kuno:1999jp,Kitano:2002mt}
\begin{align}
	\mathcal{L}_{\mathrm{eff}}= & -\frac{4 G_{\mathrm{F}}}{\sqrt{2}}\left(m_\mu A_R \bar{\mu} \sigma^{\mu \nu} P_L e F_{\mu \nu}+m_\mu A_L \bar{\mu} \sigma^{\mu \nu} P_R e F_{\mu \nu}+\text { h.c. }\right) \notag\\
	& -\frac{G_{\mathrm{F}}}{\sqrt{2}} \sum_{q=u, d, s}\bigg[\left(g_{L S}^q \bar{e} P_R \mu+g_{R S}^q \bar{e} P_L \mu\right) \bar{q} q+\left(g_{L P}^q \bar{e} P_R \mu+g_{R P}^q \bar{e} P_L \mu\right) \bar{q} \gamma_5 q\notag\\
	& +\left(g_{L V}^q \bar{e} \gamma^\mu P_{L \mu} \mu+g_{R V}^q \bar{e} \gamma^\mu P_R \mu\right) \bar{q} \gamma_\mu q+\left(g_{L A}^q \bar{e} \gamma^\mu P_{L \mu} \mu+g_{R A}^q \bar{e} \gamma^\mu P_R \mu\right) \bar{q} \gamma_\mu \gamma_5 q \notag\\
	&+\frac{1}{2}\left(g_{L T}^q \bar{e} \sigma^{\mu \nu} P_R \mu+g_{R T}^q \bar{e} \sigma^{\mu \nu} P_L \mu\right) \bar{q} \sigma_{\mu \nu} q+\text { h.c. }\bigg] \label{eft-mue}~.
\end{align}
The spin-independent conversion rate at the above basis can be written as
\begin{align}
	\Gamma_{\text {conv.}}
	&=  2 G_{\mathrm{F}}^2\left|A_R D+\tilde{g}_{L S}^p S^{(p)}+\tilde{g}_{L S}^n S^{(n)}+\tilde{g}_{L V}^p V^{(p)}+\tilde{g}_{L V}^n V^{(n)}\right|^2 \notag\\
	& +2 G_{\mathrm{F}}^2\left|A_L D+\tilde{g}_{R S}^p S^{(p)}+\tilde{g}_{R S}^n S^{(n)}+\tilde{g}_{R V}^p V^{(p)}+\tilde{g}_{R V}^n V^{(n)}\right|^2~,
\end{align}
where the dimensionless coefficients $\tilde{g}_{XS/XV}^{(n/p)}$ with $X=L,R$ are defined as 
\begin{equation}
	\begin{array}{ll}
		\tilde{g}_{L S}^p=G_S^{p,u} g_{L S}^{u}+G_S^{p,d} g_{L S}^{d}+G_S^{p,s} g_{L S}^{s}~, 
		& \tilde{g}_{R S}^p=G_S^{p,u} g_{R S}^{u}+G_S^{p,d} g_{R S}^{d}+G_S^{p,s} g_{R S}^{s}~,\\
		\tilde{g}_{L V}^p=G_V^{p,u} g_{L V}^u+G_V^{p,d}g_{L V}^d +G_V^{p,s}g_{L V}^s ~,
		&\tilde{g}_{R V}^p=G_V^{p,u} g_{R V}^u+G_V^{p,d}g_{R V}^d +G_V^{p,s}g_{R V}^s ~,\\
		\tilde{g}_{L S}^n=G_S^{ n,u} g_{L S}^{u}+G_S^{ n,d} g_{L S}^{d}+G_S^{ n,s} g_{L S}^{s}~, 
		&\tilde{g}_{R S}^n=G_S^{ n,u} g_{RS}^{u}+G_S^{ n,d} g_{R S}^{d}+G_S^{ n,s} g_{R S}^{s}~,\\
		\tilde{g}_{L V}^n=G_V^{n,u} g_{L V}^u+G_V^{n,d}g_{L V}^d+G_V^{n,s}g_{L V}^s~,
		& \tilde{g}_{R V}^n=G_V^{n,u} g_{R V}^u+G_V^{n,d}g_{R V}^d+G_V^{n,s}g_{R V}^s~.
	\end{array}
\end{equation}
The values of the $G$ factors are given in Eq. (\ref{GV}-\ref{GS2}), and the relations between the coefficients in different bases are shown as
\begin{align*}
	& A_{R}=-\dfrac{C_{e\gamma,12}^{*}}{2\sqrt{2}G_{F}m_{\mu}}~,&&
	A_{L}=-\dfrac{C_{e\gamma,21}}{2\sqrt{2}G_{F}m_{\mu}}~,\\
	&g_{LS}^{u}=-\dfrac{1}{\sqrt{2}G_{F}}[\mathcal{C}_{eu,1211}^{S,RR}+\mathcal{C}_{eu,1211}^{S,RL}]~,&&
	g_{LP}^{u}=-\dfrac{1}{\sqrt{2}G_{F}}[\mathcal{C}_{eu,1211}^{S,RR}-\mathcal{C}_{eu,1211}^{S,RL}]~, \\
	&g_{RS}^{u}=-\dfrac{1}{\sqrt{2}G_{F}}[\mathcal{C}_{eu,2111}^{S,RR*}+\mathcal{C}_{eu,2111}^{S,RL*}]~,&&
	g_{RP}^{u}=-\dfrac{1}{\sqrt{2}G_{F}}[-\mathcal{C}_{eu,2111}^{S,RR*}+\mathcal{C}_{eu,2111}^{S,RL*}]~,\\
	&g_{LV}^{u}=-\dfrac{1}{\sqrt{2}G_{F}}[\mathcal{C}_{eu,1211}^{V,LL}+\mathcal{C}_{eu,1211}^{V,LR}]~,&&
	g_{LA}^{u}=-\dfrac{1}{\sqrt{2}G_{F}}[-\mathcal{C}_{eu,1211}^{V,LL}+\mathcal{C}_{eu,1211}^{V,LR}]~, \\
	&g_{RV}^{u}=-\dfrac{1}{\sqrt{2}G_{F}}[\mathcal{C}_{eu,1211}^{V,RR}+\mathcal{C}_{ue,1112}^{V,LR}]~,&&
	g_{RA}^{u}=-\dfrac{1}{\sqrt{2}G_{F}}[\mathcal{C}_{eu,1211}^{V,RR}-\mathcal{C}_{ue,1112}^{V,LR}]~,\\
	&g_{LT}^{u}=-\dfrac{2\sqrt{2}}{G_{F}}\mathcal{C}_{eu,1211}^{T,RR}~,&&
	g_{RT}^{u}=-\dfrac{2\sqrt{2}}{G_{F}}\mathcal{C}_{eu,2111}^{T,RR*}~,\\
	&g_{LS}^{d}=-\dfrac{1}{\sqrt{2}G_{F}}[\mathcal{C}_{ed,1211}^{S,RR}+\mathcal{C}_{ed,1211}^{S,RL}]~,&&
	g_{LP}^{d}=-\dfrac{1}{\sqrt{2}G_{F}}[\mathcal{C}_{ed,1211}^{S,RR}-\mathcal{C}_{ed,1211}^{S,RL}]~,\\
	&g_{RS}^{d}=-\dfrac{1}{\sqrt{2}G_{F}}[\mathcal{C}_{ed,2111}^{S,RR*}+\mathcal{C}_{ed,2111}^{S,RL*}]~,&&
	g_{RP}^{d}=-\dfrac{1}{\sqrt{2}G_{F}}[-\mathcal{C}_{ed,2111}^{S,RR*}+\mathcal{C}_{ed,2111}^{S,RL*}]~,\\
	&g_{LV}^{d}=-\dfrac{1}{\sqrt{2}G_{F}}[\mathcal{C}_{ed,1211}^{V,LL}+\mathcal{C}_{ed,1211}^{V,LR}]~,&&
	g_{LA}^{d}=-\dfrac{1}{\sqrt{2}G_{F}}[-\mathcal{C}_{ed,1211}^{V,LL}+\mathcal{C}_{ed,1211}^{V,LR}]~,\\
	&g_{RV}^{d}=-\dfrac{1}{\sqrt{2}G_{F}}[\mathcal{C}_{ed,1211}^{V,RR}+\mathcal{C}_{de,1112}^{V,LR}]~,&&
	g_{RA}^{d}=-\dfrac{1}{\sqrt{2}G_{F}}[\mathcal{C}_{ed,1211}^{V,RR}-\mathcal{C}_{de,1112}^{V,LR}]~,\\
	&g_{LT}^{d}=-\dfrac{2\sqrt{2}}{G_{F}}\mathcal{C}_{ed,1211}^{T,RR}~,&&
	g_{RT}^{d}=-\dfrac{2\sqrt{2}}{G_{F}}\mathcal{C}_{ed,2111}^{T,RR*}~,\\
	&g_{LS}^{s}=-\dfrac{1}{\sqrt{2}G_{F}}[\mathcal{C}_{ed,1222}^{S,RR}+\mathcal{C}_{ed,1222}^{S,RL}]~,&&
	g_{LP}^{s}=-\dfrac{1}{\sqrt{2}G_{F}}[\mathcal{C}_{ed,1222}^{S,RR}-\mathcal{C}_{ed,1222}^{S,RL}]~,\\
	&g_{RS}^{s}=-\dfrac{1}{\sqrt{2}G_{F}}[\mathcal{C}_{ed,2122}^{S,RR*}+\mathcal{C}_{ed,2122}^{S,RL*}]~,&&
	g_{RP}^{s}=-\dfrac{1}{\sqrt{2}G_{F}}[-\mathcal{C}_{ed,2122}^{S,RR*}+\mathcal{C}_{ed,2122}^{S,RL*}]~,\\
	&g_{LV}^{s}=-\dfrac{1}{\sqrt{2}G_{F}}[\mathcal{C}_{ed,1222}^{V,LL}+\mathcal{C}_{ed,1222}^{V,LR}]~,&&
	g_{LA}^{s}=-\dfrac{1}{\sqrt{2}G_{F}}[-\mathcal{C}_{ed,1222}^{V,LL}+\mathcal{C}_{ed,1222}^{V,LR}]~,\\
	&g_{RV}^{s}=-\dfrac{1}{\sqrt{2}G_{F}}[\mathcal{C}_{ed,1222}^{V,RR}+\mathcal{C}_{de,2212}^{V,LR}]~,&&
	g_{RA}^{s}=-\dfrac{1}{\sqrt{2}G_{F}}[\mathcal{C}_{ed,1222}^{V,RR}-\mathcal{C}_{de,2212}^{V,LR}]~,\\
	&g_{LT}^{s}=-\dfrac{2\sqrt{2}}{G_{F}}\mathcal{C}_{ed,1222}^{T,RR}~,&&
	g_{RT}^{s}=-\dfrac{2\sqrt{2}}{G_{F}}\mathcal{C}_{ed,2122}^{T,RR*}~.
\end{align*}

The preceding discussion yields a general formula for the $\mu$-$e$ conversion. To enable a more concise and focused analysis, this formula will now be explicitly presented below within the $\slashed{R}$-SUSY model, allowing subsequent sections to apply and analyze it directly.

\vspace*{-0.4\baselineskip}
\subsection{The \texorpdfstring{$\mu$-$e$}{} conversion formula for \texorpdfstring{$R$}{}-parity violating SUSY model}
In the $R$-parity violating SUSY model, there are no contributions to the scalar type and the tensor type four-fermion operators. This follows directly from the matching conditions from the UV to the SMEFT, which involve only the vector type $qq\ell\ell$ operators. The $\mu$-$e$ conversion rate related to the spin-independent interaction can be expressed as
\begin{align}
\Gamma_{\text {conv.}}
&= 
\big|
D C_{e\gamma,12}^{*}/(2m_{\mu})+(2\mathcal{C}_{eu,1211}^{V,LL}+2\mathcal{C}_{eu,1211}^{V,LR}+\mathcal{C}_{ed,1211}^{V,LL}+\mathcal{C}_{ed,1211}^{V,LR})V^{(p)}\notag\\
&+(\mathcal{C}_{eu,1211}^{V,LL}+\mathcal{C}_{eu,1211}^{V,LR}+2\mathcal{C}_{ed,1211}^{V,LL}+2\mathcal{C}_{ed,1211}^{V,LR})V^{(n)}\big|^{{2}}\notag\\
&+\big|D C_{e\gamma,21}/(2m_{\mu})
+(2\mathcal{C}_{eu,1211}^{V,RR}+2\mathcal{C}_{ue,1112}^{V,LR}+\mathcal{C}_{ed,1211}^{V,RR}+\mathcal{C}_{de,1112}^{V,LR})V^{(p)}\notag\\
&+(\mathcal{C}_{eu,1211}^{V,RR}+\mathcal{C}_{ue,1112}^{V,LR}+2\mathcal{C}_{ed,1211}^{V,RR}+2\mathcal{C}_{de,1112}^{V,LR})V^{(n)}\big|^{{2}}~,
\end{align}
with the overlap integral values have already been shown in Tab.~\ref{values-isotopes}. Using this formula and the numerical RG effects shown in Tab.~\ref{RGE}, one can give constraints on the parameters based on experimental limits. To investigate the ability to examine models using $\mu$-$e$ conversion, we will discuss the formulas for muon decays $\mu\to e\gamma$ and $\mu\to3e$ to complement the analysis of muon cLFV in the following section.

\section{Contrasts from \texorpdfstring{$\bm{\mu\to e\gamma$}}{} and \texorpdfstring{\bm{$\mu\to 3e$}}{}}
The branching ratio of the muon radiative decay $\mu\to e \gamma$ can be expressed in terms of the Wilson coefficients of the dimension-5 LEFT operators~\cite{Lavoura:2003xp}
\begin{align}
\text{Br}(\mu\to e \gamma)
&=\tau_{\mu}\times\dfrac{(m_{\mu}^{2}-m_{e}^{2})^{3}}{4\pi m_{\mu}^{3}}(|\mathcal{C}_{e\gamma,12}|^{2}+|\mathcal{C}_{e\gamma,21}|^{2})~,
\end{align}
where $\tau_{\mu}=192\pi^{3}/(G_{F}^{2}m_{\mu}^{5})$ is the life-time of muon.
The branching ratio of $\mu\to 3 e$ decay can be described as \cite{deGouvea:2000cf,Crivellin:2017rmk,Ardu:2024bua}
\begin{align}
\text{Br}(\mu\to 3 e)
=&\dfrac{1}{64G_{F}^{2}}\left(\left|\mathcal{C}_{ee,1121}^{S,RR}\right|^{2}+\left|\mathcal{C}_{ee,1112}^{S,RR}\right|^{2}\right)\notag\\
+&\dfrac{\alpha_{\text{em}}}{\pi G_{F}^{2}m_{\mu}^{2}}
\bigg(\ln\dfrac{m_{\mu}^{2}}{m_{e}^{2}}-\dfrac{17}{4}\bigg)\left(\left|\mathcal{C}_{e\gamma,12}\right|^{2}+\left|\mathcal{C}_{e\gamma,21}\right|^{2}\right)\notag\\
+&\dfrac{1}{8G_{F}^{2}}
\bigg(2\bigg|\mathcal{C}_{ee,1112}^{V,RR}+\dfrac{4e}{m_{\mu}}\mathcal{C}_{e\gamma,12}^{*}\bigg|^{2}
+2\bigg|\mathcal{C}_{ee,1112}^{V,LL}+\dfrac{4e}{m_{\mu}}\mathcal{C}_{e\gamma,21}\bigg|^{2}\notag\\
&\quad\quad+\bigg|\mathcal{C}_{ee,1112}^{V,LR}+\dfrac{4e}{m_{\mu}}\mathcal{C}_{e\gamma,12}^{*}\bigg|^{2}
+\left|\mathcal{C}_{ee,1211}^{V,LR}+\dfrac{4e}{m_{\mu}}\mathcal{C}_{e\gamma,21}\right|^{2}\bigg)~,
\end{align}
where $e$ is the coupling strength of electromagnetic interaction, and $\alpha_{\text{em}}=e^{2}/4\pi$ is the fine structure constant. If only contributions from dim-5 operators are considered, then the branch ratio can be simplified as
\begin{align}
\text{Br}(\mu\to 3 e)=\dfrac{16\pi \alpha_{\text{em}}}{G_{F}^{2}m_{\mu}^{2}}\left(\ln\dfrac{m_{\mu}^{2}}{m_{e}^{2}}-\dfrac{11}{4}\right)\left(\left|\mathcal{C}_{e\gamma,12}\right|^{2}+\left|\mathcal{C}_{e\gamma,21}\right|^{2}\right)~,
\end{align}
and as the previous literature~\cite{Petcov:1977ab,Ilakovac:1994kj,Hisano:1995cp,Hisano:1998fj,Bolton:2022lrg} have been calculated and clarified, one can obtain
\begin{align}
\dfrac{\text{Br}(\mu\to 3 e)}{\text{Br}(\mu\to e \gamma)}
=\dfrac{\alpha_{\text{em}}}{3\pi}\left(\ln\dfrac{m_{\mu}^{2}}{m_{e}^{2}}-\dfrac{11}{4}\right)\simeq \dfrac{1}{160}~.
\end{align}

Here is a simple example to illustrate the ability of different muon cLFV processes to constrain the parameters. By improving the results in \cite{Delzanno:2024ooj}, one can obtain the updated upper bound from current $\mu\to e\gamma$ decay experimental constraints
\begin{align}
\left|\mathcal{C}_{e\gamma,12}\right|^{2}+\left|\mathcal{C}_{e\gamma,21}\right|^{2}
<4.812\times 10^{-22} ~\text{[TeV]}^{-2}~,\quad(\text{MEG-II},~\mu\to e\gamma~\text{decay})~.
\end{align}
This upper limit can be improved to $\sim2\times10^{-22}~[\text{TeV}]^{-2}$ in the near future, assuming no signal is observed. If the $\mu N\to e N$ conversion and $\mu\to 3e$ decay are also dominated by the operator $\mathcal{O}_{e\gamma}$, then one can obtain the upper limits on the Wilson coefficients from the current experimental searching results
\begin{align}
&<4.10\times 10^{-20} ~\text{[TeV]}^{-2}~,\quad(\text{SINDRUM-II},~\mu \text{Ti}\to e \text{Ti})~;\\
\left|\mathcal{C}_{e\gamma,12}\right|^{2}+\left|\mathcal{C}_{e\gamma,21}\right|^{2}
&<1.05\times 10^{-19} ~\text{[TeV]}^{-2}~,\quad(\text{SINDRUM-II},~\mu \text{Au}\to e \text{Au})~;\\
&<1.910\times 10^{-18} ~\text{[TeV]}^{-2}~,\quad(\text{SINDRUM},~\mu\to 3e~\text{decay})~.
\end{align}
According to the sensitivities shown in Tab.~\ref{exp}, the future Mu3e experiment, which aims to searching for the $\mu\to 3e$ decay, can improve the limits to $10^{-21}~\text{[TeV]}^{-2}$ in Phase I and reach $10^{-22}~\text{[TeV]}^{-2}$ in Phase II, if no such decay is observed. The constraint can also be improved to $10^{-22}~\text{[TeV]}^{-2}$  by COMET Phase I and to $10^{-23}~\text{[TeV]}^{-2}$ by Mu2e Run I. The COMET Phase II and Mu2e Run II can further enhance the sensitivity about $10^{-24}~\text{[TeV]}^{-2}$. 

As one can easily infer, the current experimental limit on $\mu\to e\gamma$ decay can provide much more stringent constraints on the Wilson coefficients than the $\mu N\to e N$ conversion and $\mu\to 3 e$ decay searching experiments when the dim-5 operators dominate the contributions. However, in the near future, $\mu N\to eN$ conversion experiments are also expected to reach the same level of sensitivity and further improve this limit.

\section{The numerical results and discussion}
In this section, we focus on numerical results for the combinations of $\lambda$ and $\lambda^{\prime}$ that can induce the cLFV processes. Using experimental limits, one can derive the constraints on the combinations of coupling parameters in the form of $|\lambda^{(\prime)}_{ijk}\lambda^{(\prime)*}_{prs}|$, expressed in units of $[m_{\tilde{f}}^2(\text{TeV}^{-2})]$. For simplicity, we assume only one such combination is nonzero at a time.

The upper limits on the parameters of $\lambda^{\prime}$ term are shown in Tab.~\ref{lambdap-results} and Tab.~\ref{lambdap-results2}, while the limits on $\lambda$ parameters are shown in Tab.~\ref{lambda-results}. We show the limits from $\mu$-$e$ conversion in the second (with RG effects) and the third column (without RG effects) with the current experimental constraints on Ti and Au conversion branching ratio from SINDRUM II and the future sensitivity on Al from COMET phase I which is around $3\times10^{-15}$ as already shown in Tab.~\ref{exp}. The limits from muon decay $\mu\to e\gamma$ are shown in the fourth and fifth columns with the current (future) experimental constraints from MEG-II, while those from $\mu\to 3e$ are in the last two columns with the current (future) constraints from SINDRUM (Mu3e phase I).

\subsection{Results for $\lambda^{\prime}$ couplings}

\begin{table}[tb]
\renewcommand{\arraystretch}{1.2}
\resizebox{\textwidth}{!}{
\begin{tabular}{|c||c|c||c|c||c|c|}
\hline\hline
\multirow{3}*{Para.} & Limits  & Limits & Limits  &Limits& Limits& Limits\\
& w/ RG effects &  w/o RG effects & w RG effects &w/o RG effects & w RG effects &w/o RG effects\\
& $\mu-e$ conv. & $\mu-e$ conv. & $\mu\to e \gamma$& $\mu\to e \gamma$ & $\mu\to 3e$ & $\mu\to 3e$ \\
\hline
\multirow{3}*{$|\lambda_{211}^{\prime}\lambda_{111}^{\prime *}|$}
& $9.32\times 10^{-5}$ (Ti) & $4.03\times 10^{-5}$ (Ti) & & & &\\
& $3.44\times 10^{-5}$ (Au) & $2.35\times 10^{-5}$ (Au)  
&$9.48\times 10^{-4}$ & $8.63\times 10^{-4}$  
&$1.14\times 10^{-3}$ & $1.33\times 10^{-3}$\\
& $2.72\times 10^{-5}$ (Al) & $5.32\times 10^{-6}$ (Al) 
& ($6.00\times10^{-4}$)& ($5.46\times10^{-4}$)
& ($5.10\times10^{-5}$)& ($5.96\times10^{-5}$)\\
\hline
\multirow{3}*{$|\lambda_{212}^{\prime}\lambda_{112}^{\prime *}|$}
& $4.22\times 10^{-6}$ (Ti)& $4.57\times 10^{-6}$ (Ti) & & & &\\
&  $4.35\times 10^{-6}$ (Au)  & $4.71\times 10^{-6}$ (Au) 
&$9.48\times 10^{-4}$  &$8.63\times 10^{-4}$ 
&$1.25\times 10^{-3}$  & $1.45\times 10^{-3}$\\
& $3.94\times 10^{-7}$ (Al)& $4.25\times 10^{-7}$ (Al) 
& ($6.00\times10^{-4}$)& ($5.46\times10^{-4}$)
& ($5.57\times10^{-5}$)& ($6.49\times10^{-5}$)\\
\hline
\multirow{3}*{$|\lambda_{213}^{\prime}\lambda_{113}^{\prime *}|$}
& $4.22\times 10^{-6}$ (Ti)& $4.57\times 10^{-6}$ (Ti) & & & &\\
& $4.35\times 10^{-6}$ (Au)& $4.71\times 10^{-6}$ (Au)
& $9.48\times 10^{-4}$&$8.63\times 10^{-4}$ 
&$1.41\times 10^{-3}$ & $1.63\times 10^{-3}$\\
& $3.94\times 10^{-7}$ (Al)  & $4.25\times 10^{-7}$ (Al) 
& ($6.00\times10^{-4}$)& ($5.46\times10^{-4}$)
&  ($6.30\times10^{-5}$)& ($7.30\times10^{-5}$)\\
\hline
\multirow{3}*{$|\lambda_{221}^{\prime}\lambda_{121}^{\prime *}|$}
& $4.21\times 10^{-6}$ (Ti)& $4.31\times 10^{-6}$ (Ti) & & & &\\
&$4.02\times 10^{-6}$ (Au) &$4.10\times 10^{-6}$ (Au) 
& $9.48\times 10^{-4}$& $8.63\times 10^{-4}$ 
&$1.68\times 10^{-3}$ & $1.93\times 10^{-3}$\\
&$4.05\times 10^{-7}$ (Al) & $4.14\times 10^{-7}$ (Al) 
& ($6.00\times10^{-4}$)& ($5.46\times10^{-4}$)
&  ($7.50\times10^{-5}$)& ($8.64\times10^{-5}$)\\
\hline
\multirow{3}*{$|\lambda_{231}^{\prime}\lambda_{131}^{\prime *}|$}
& $4.12\times 10^{-6}$ (Ti)& $4.10\times 10^{-6}$ (Ti) & & & &\\
& $3.95\times 10^{-6}$ (Au) & $3.92\times 10^{-6}$ (Au)
& $1.09\times 10^{-3}$& $9.92\times 10^{-4}$ 
&$2.92\times 10^{-3}$ & $3.30\times 10^{-3}$\\
& $3.96\times 10^{-7}$ (Al) & $3.94\times 10^{-7}$ (Al) 
&($6.89\times10^{-4}$) & ($6.27\times10^{-4}$)
&($1.31\times10^{-4}$) & ($1.47\times10^{-4}$) \\
\hline\hline
\end{tabular}}
\caption{The upper limits with and without RG running effects on the combinations of $\lambda^{\prime}$ couplings from cLFV processes $\mu$-$e$ conversion, $\mu\to e\gamma $ and $\mu\to 3 e$. The units of these upper limits are $[m_{\tilde{q}}^{2} ~(\text{TeV}^{-2})]$. }\label{lambdap-results}
\end{table}

\begin{table}[t]
\resizebox{\textwidth}{!}{
\renewcommand{\arraystretch}{1.17}
\begin{tabular}{|c||c|c||c|c||c|c|}
\hline\hline
\multirow{3}*{Para.} &Limits  & Limits & Limits  &Limits & Limits  &Limits\\
& w/ RG effects &  w/o RG effects & w/ RG effects &  w/o RG effects & w/ RG effects &  w/o RG effects\\
& $\mu-e$ conv. & $\mu-e$ conv. & $\mu\to e \gamma$  & $\mu\to e \gamma$ &  $\mu\to 3e $ & $\mu\to 3e $\\
\hline
\multirow{3}*{$|\lambda_{222}^{\prime}\lambda_{122}^{\prime *}|$}
& $9.94\times 10^{-5}$ (Ti) &  $8.62\times 10^{-5}$ (Ti) & & & &\\
& $1.00\times 10^{-4}$ (Au) &  $8.89\times 10^{-5}$ (Au) 
& $9.48\times 10^{-4}$&$8.63\times 10^{-4}$ 
&$1.91\times 10^{-3}$ & $2.19\times 10^{-3}$\\
& $9.33\times 10^{-6}$ (Al) & $8.03\times 10^{-6}$ (Al) 
& ($6.00\times10^{-4}$)& ($5.46\times10^{-4}$)
&  ($8.56\times10^{-5}$)& ($9.81\times10^{-5}$)\\
\hline
\multirow{3}*{$|\lambda_{223}^{\prime}\lambda_{123}^{\prime *}|$}
& $9.94\times 10^{-5}$ (Ti) &  $8.62\times 10^{-5}$ (Ti) & & & &\\
& $1.00\times 10^{-4}$ (Au) &  $8.89\times 10^{-5}$ (Au) 
&$9.48\times 10^{-4}$ & $ 8.63\times 10^{-4}$ 
&$2.32\times 10^{-3}$ & $2.64\times 10^{-3}$\\
& $9.34\times 10^{-6}$ (Al) & $8.03\times 10^{-6}$ (Al) 
& ($6.00\times10^{-4}$)& ($5.46\times10^{-4}$)
&  ($1.04\times10^{-4}$)& ($1.18\times10^{-4}$)\\
\hline
\multirow{3}*{$|\lambda_{22k}^{\prime}\lambda_{11k}^{\prime *}|$}
& $1.81\times 10^{-5}$ (Ti)&  $1.98\times 10^{-5}$ (Ti) & & & &\\
& $1.87\times 10^{-5}$ (Au) & $2.05\times 10^{-5}$ (Au) 
& / & /
& / & /\\
& $1.69\times 10^{-6}$ (Al) &  $1.85\times 10^{-6}$ (Al) & & & &\\
\hline
\multirow{3}*{$|\lambda_{21k}^{\prime}\lambda_{12k}^{\prime *}|$}
& $1.81\times 10^{-5}$ (Ti)& $1.98\times 10^{-5}$ (Ti) & & & &\\
& $1.87\times 10^{-5}$ (Au) & $2.05\times 10^{-5}$ (Au) 
& / & / 
& / & /\\
& $1.69\times 10^{-6}$ (Al) & $1.85\times 10^{-6}$ (Al) & & & &\\
\hline
\multirow{3}*{$|\lambda_{23k}^{\prime}\lambda_{11k}^{\prime *}|$}
& $3.00\times 10^{-3}$ (Ti)&  $4.16\times 10^{-3}$ (Ti) & & & &\\
& $3.01\times 10^{-3}$ (Au) & $4.29\times 10^{-3}$ (Au) 
& / & /
& / &/\\
& $2.82\times 10^{-4}$ (Al) & $3.87\times 10^{-4}$ (Al) & & & &\\
\hline
\multirow{3}*{$|\lambda_{21k}^{\prime}\lambda_{13k}^{\prime *}|$}
& $3.00\times 10^{-3}$ (Ti)&  $4.16\times 10^{-3}$ (Ti) & & & &\\
& $3.01\times 10^{-3}$ (Au) & $4.29\times 10^{-3}$ (Au) 
& / &/ 
& / &/\\
& $2.82\times 10^{-4}$ (Al) & $3.87\times 10^{-4}$ (Al)& & & &\\
\hline
\multirow{3}*{$|\lambda_{232}^{\prime}\lambda_{132}^{\prime *}|$}
& $2.76\times 10^{-4}$ (Ti)& $7.72\times 10^{-4}$ (Ti) &  & & &\\
& $2.07\times 10^{-4}$ (Au) & $8.76\times 10^{-4}$ (Au) 
&$1.09\times 10^{-3}$ & $9.92\times 10 ^{-4}$ 
& $3.72\times 10^{-3}$& $4.13\times 10^{-3}$\\
& $3.00\times 10^{-5}$ (Al)  & $6.97\times 10^{-5}$ (Al)
&($6.89\times10^{-4}$) & ($6.27\times10^{-4}$)
&($1.66\times10^{-4}$) &($1.85\times10^{-4}$) \\
\hline
\multirow{3}*{$|\lambda_{233}^{\prime}\lambda_{133}^{\prime *}|$}
& $2.63\times 10^{-4}$ (Ti)& $8.90\times 10^{-4}$ (Ti) & & & &\\
& $2.01\times 10^{-4}$ (Au) & $1.01\times 10^{-3}$ (Au)
& $1.09\times 10^{-3}$ & $9.92\times 10 ^{-4}$  
& $5.66\times 10^{-3}$& $6.05\times 10^{-3}$\\
& $2.84\times 10^{-5}$ (Al) & $8.03\times 10^{-5}$ (Al)
&($6.89\times10^{-4}$) & ($6.27\times10^{-4}$)
 &($2.53\times10^{-4}$) &($2.71\times10^{-4}$) \\
\hline
\multirow{3}*{$|\lambda_{23k}^{\prime}\lambda_{12k}^{\prime *}|$}
& $3.88\times 10^{-3}$ (Ti)& $1.81\times 10^{-2}$ (Ti)  &  & & &\\
& $3.43\times 10^{-3}$ (Au) &  $1.86\times 10^{-2}$ (Au)
& /  & / & / &/\\
& $3.86\times 10^{-4}$ (Al) & $1.68\times 10^{-3}$ (Al) & & & &\\
\hline
\multirow{3}*{$|\lambda_{22k}^{\prime}\lambda_{13k}^{\prime *}|$}
& $3.88\times 10^{-3}$ (Ti)& $1.81\times 10^{-2}$ (Ti)  &  & & &\\
& $3.43\times 10^{-3}$ (Au) &  $1.86\times 10^{-2}$ (Au)
& /  & / & / &/\\
& $3.86\times 10^{-4}$ (Al) & $1.68\times 10^{-3}$ (Al) & & & &\\
\hline\hline
\end{tabular}}
\caption{The upper limits with and without RG effects on the combinations of $\lambda^{\prime}$ couplings from cLFV processes $\mu$-$e$ conversion, $\mu\to e\gamma $ and $\mu\to 3 e$. The units of these upper limits are $[m_{\tilde{q}}^{2} ~(\text{TeV}^{-2})]$. }\label{lambdap-results2}
\end{table}

In Tab.~\ref{lambdap-results}, the limits from $\mu$-$e$ conversion are much stronger than those from $\mu\to e\gamma$ and $\mu\to 3e$, since the combinations can give tree level contributions to $\mu$-$e$ conversion. These results are in strong agreement with the previous results in \cite{Barbier:2004ez}. Accounting for RG effects can improve the constraints by no more than 10\% in most cases, except for the combination $|\lambda_{211}^{\prime}\lambda_{111}^{\prime *}|$. The combination $|\lambda_{211}^{\prime}\lambda_{111}^{\prime *}|$ can contribute to $[\mathcal{C}_{\ell q}^{(1,3)}]_{1211}$ and $[\mathcal{C}_{\ell d}]_{1211}$ which would lead to a weaker upper bound than the other combinations due to cancellations among different contributions.
If no RG effects are considered, the conversion rate becomes
\begin{align}
\Gamma_{\text{conv.}}\simeq|\lambda_{211}^{\prime}\lambda_{111}^{\prime*}|^{2} |(V^{(p)}-V^{(n)})(1/2m_{\tilde{q}}^{2})|^{2}~,
\end{align}
one can find that $V^{(p)}$ and $V^{(n)}$ have similar values in those isotopes, inducing the cancellation. The branching ratio is proportional to $\text{Br}\propto |V^{(p)}-V^{(n)}|^{2}/\Gamma_{\text{cap.}}$, which has a larger value for Au results, giving about 2 times stronger upper limits than for Ti.
Including the RG effect can further enhance these cancellations and weaken the constraint by 30\%-50\%. 

Note that $\lambda_{111}^{\prime}$ can give a short-range contribution to neutrinoless double beta decay via the exchange of neutralino and gluino \cite{Hirsch:1995ek,Cirigliano:2004tc}. The inverse half-life $T_{1/2}^{-1}$ upper limits from KamLAND-Zen~\cite{KamLAND-Zen:2024eml} and GERDA~\cite{GERDA:2020xhi} can give constraints $ \lambda_{111}^{\prime 2} / m_{\tilde{q}}^{4}  m_{\tilde{g}} < 3\sim 4 \times 10^{-4}~ \text{TeV}^{-5} $, which implies $\lambda_{111}^{\prime}<10^{-2}$ for typical TeV scale of squark and gluino masses. The constraints may be weakened if standard light neutrino exchange and SUSY particle exchange occur simultaneously, since cancellations between the contributions are possible~\cite{Agostini:2022bjh,Chen:2024ctu}. The literature \cite{Bolton:2021hje} has shown that a light neutralino exchange with mass $m_{\tilde{\tilde{\chi}}^{0}}$ below the GeV scale leads to a much stronger bound, $\lambda_{111}^{\prime}<10^{-4}$, comparable to the limit from the non-observation of $\tilde{\chi}_{1}^{0}$ decay in Super-Kamiokande~\cite{Candia:2021bsl}.

In Tab.~\ref{lambdap-results2}, we list upper limits on other 10 combinations of $\lambda^{\prime}$. Certain of these combinations lack limits from $\mu\to e\gamma$ and $\mu\to 3e$ because of the GIM suppression, and as a result they have been largely neglected in previous studies. However, when CKM matrix elements are inserted, these combinations can in fact produce dominant tree-level contributions to $\mu$-$e$ conversion. For $\lambda_{232}^{\prime} \lambda_{132}^{\prime}$ and $\lambda_{233}^{\prime} \lambda_{133}^{\prime}$, the tree level contributions to the conversion are highly suppressed by the CKM matrice element $V_{\text{CKM},t1}$, and the dominant contributions are from the one-loop level. The RG effects can give 10\%-30\% influence on the results in most combinations, while can give 80\% improvement on the upper limits for $\lambda_{23k}^{\prime} \lambda_{12k}^{\prime *}$ and $\lambda_{13k}^{\prime} \lambda_{22k}^{\prime *}$ combinations. 

Note that the uncertainties about $2\%\sim 5\%$ in the nuclear distributions discussed in Sec. \ref{3.1} do not significantly affect the current upper limits or future sensitivities for the low $Z$ isotopes $^{48}$Ti and $^{27}$Al. Likewise, the large RG effects in specific combinations remain robust against nuclear uncertainties, even when considering the high $Z$ isotope $^{197}$Au with a $10\%$ uncertainty.

One should also pay attention to the constraints from neutrino mass as the $R$-parity violating SUSY model can be an origin of Majorana neutrino mass via one-loop with  $\lambda^{\prime}$ or $\lambda$ combinations~\cite{Hall:1983id,Babu:1989px}. The realization from $\lambda^{\prime}$ term can be expressed as
\begin{align}
M_{\nu,ij}\simeq
\dfrac{3}{8\pi^2}
\bigg[&\lambda_{i11}^{\prime}\lambda_{j11}^{\prime}\dfrac{m_{d}^{2}}{\tilde{m}}
+\lambda_{i22}^{\prime}\lambda_{j22}^{\prime}\dfrac{m_{s}^{2}}{\tilde{m}}
+\lambda_{i33}^{\prime}\lambda_{j33}^{\prime}\dfrac{m_{b}^{2}}{\tilde{m}}
+(\lambda_{i12}^{\prime}\lambda_{j21}^{\prime}+\lambda_{i21}^{\prime}\lambda_{j12}^{\prime})\dfrac{m_{d}m_{s}}{\tilde{m}}\notag\\
+&(\lambda_{i13}^{\prime}\lambda_{j31}^{\prime}+\lambda_{i31}^{\prime}\lambda_{j13}^{\prime})\dfrac{m_{d}m_{b}}{\tilde{m}}
+(\lambda_{i23}^{\prime}\lambda_{j32}^{\prime}+\lambda_{i32}^{\prime}\lambda_{j23}^{\prime})\dfrac{m_{s}m_{b}}{\tilde{m}}
\bigg]~.\label{vmass-1}
\end{align}
With the constraints on the sum of neutrino masses $\sum_{i}m_{i}\lesssim 0.13~\text{eV}$ \cite{ACT:2023kun}, one can derive the upper limits for the combination $|\lambda_{133}^{\prime}\lambda_{233}^{\prime}|<\mathcal{O}(10^{-7}\sim10^{-8})~\times(\tilde{m}~\text{TeV}^{-1})$ with $\tilde{m}=m_{\tilde{q}}^{2}/(A^{d}-\mu\tan{\beta})$. The constraints will be much stronger than from the cLFV processes with a typical value for $\tilde{m}$ to be from 100 GeV to 1 TeV. The neutrino mass can also give constraints on $\lambda_{211}^{\prime}\lambda_{111}^{\prime*}$ and $\lambda_{222}^{\prime}\lambda_{122}^{\prime*}$, but the upper limits are much weaker than the $\mu$-$e$ conversion can give.

The combinations $\lambda_{22k}^{\prime} \lambda_{11k}^{\prime *}$ and $\lambda_{12k}^{\prime} \lambda_{21k}^{\prime *}$ can also contribute to the $K\to\pi\nu\bar{\nu}$ decay. The experimental branching ratio $\text{Br}=1.47\times10^{-10}$~\cite{NA62:2024pjp} implies upper constraints on these combinations  of order $\mathcal{O}(10^{-4})$~\cite{Deandrea:2004ae} for TeV scale sparticle masses, which are weaker than the bounds from $\mu$-$e$ conversion. The combinations $\lambda_{23k}^{\prime} \lambda_{12k}^{\prime *}$ and $\lambda_{13k}^{\prime} \lambda_{22k}^{\prime *}$ can contribute to $B \to X_s \nu \bar{\nu}$, but the upper limits on the combinations from current experimental results~\cite{Grossman:1995gt,Belle-II:2025bho,Fael:2025xmi} are much weaker than those from the cLFV processes.

\subsection{Results for $\lambda$ couplings}
\begin{table}[t]
\resizebox{\textwidth}{!}{
\renewcommand{\arraystretch}{1.2}
\begin{tabular}{|c||c|c||c|c||c|c|}
\hline\hline
\multirow{3}*{Para.} &Limits  & Limits & Limits  &Limits & Limits  &Limits\\
& w/ RG effects &  w/o RG effects & w/ RG effects &  w/o RG effects & w/ RG effects &  w/o RG effects\\
& $\mu-e$ conv. & $\mu-e$ conv. & $\mu\to e \gamma$  & $\mu\to e \gamma$ &  $\mu\to 3e $ & $\mu\to 3e $\\
\hline
\multirow{3}*{$|\lambda_{121}\lambda_{122}^{*}|$}
& $3.83\times10^{-4}$ (Ti) & $3.81\times10^{-4}$ (Ti) & & & & \\
& $4.34\times10^{-4}$ (Au) & $4.32\times10^{-4}$ (Au) 
& $1.42\times10^{-3}$ & $1.29\times10^{-3}$
&  $6.38\times10^{-5}$ & $6.60\times10^{-5}$\\
& $3.46\times10^{-5}$ (Al) & $3.42\times10^{-5}$ (Al) 
&  ($9.01\times10^{-4}$) & ($8.18\times10^{-4}$)
& ($2.85\times10^{-6}$) &($2.95\times10^{-6}$)\\
\hline
\multirow{3}*{$|\lambda_{131}\lambda_{132}^{*}|$}
& $4.38\times10^{-4}$ (Ti) & $4.37\times10^{-4}$ (Ti) & & & &\\
& $4.97\times10^{-4}$ (Au) & $4.94\times10^{-4}$ (Au) 
& $1.42\times10^{-3}$ & $1.29\times10^{-3}$
&  $6.38\times10^{-5}$ & $6.60\times10^{-5}$\\
& $3.96\times10^{-5}$ (Al) & $3.94\times10^{-5}$ (Al) 
&  ($9.01\times10^{-4}$) & ($8.18\times10^{-4}$)
 & ($2.85\times10^{-6}$)&($2.95\times10^{-6}$)\\
\hline
\multirow{3}*{$|\lambda_{231}\lambda_{131}^{*}|$}
& $6.22\times10^{-4}$ (Ti) & $6.20\times10^{-4}$ (Ti) & & & &\\
& $7.04\times10^{-4}$ (Au) & $7.01\times10^{-4}$ (Au) 
&  $2.85\times10^{-3}$ & $2.59\times10^{-3}$
&  $6.35\times10^{-5}$ & $6.60\times10^{-5}$\\
& $5.62\times10^{-5}$ (Al) & $5.60\times10^{-5}$ (Al) 
&  ($1.80\times10^{-3}$) & ($1.64\times10^{-3}$)
 & ($2.84\times10^{-6}$)&($2.95\times10^{-6}$)\\
\hline
\multirow{3}*{$|\lambda_{231}\lambda_{232}^{*}|$}
& $6.05\times10^{-4}$ (Ti) & $6.02\times10^{-4}$ (Ti) & & & &\\
& $6.86\times10^{-4}$ (Au) & $6.82\times10^{-4}$ (Au) 
& $1.42\times10^{-3}$ & $1.29\times10^{-3}$
&  $3.26\times10^{-3}$ & $3.26\times10^{-3}$\\
& $5.46\times10^{-5}$ (Al) & $5.43\times10^{-5}$ (Al) 
&  ($9.01\times10^{-4}$) & ($8.18\times10^{-4}$)
& ($1.47\times10^{-4}$)& ($1.46\times10^{-4}$)\\
\hline
\multirow{3}*{$|\lambda_{232}\lambda_{132}^{*}|$}
& $1.02\times10^{-3}$ (Ti) & $1.02\times10^{-3}$ (Ti) & &  & &\\
& $1.16\times10^{-3}$ (Au) & $1.15\times10^{-3}$ (Au) 
&  $2.85\times10^{-3}$ & $2.59\times10^{-3}$
&  $5.62\times10^{-3}$ & 
$5.62\times10^{-3}$\\
& $9.21\times10^{-5}$ (Al) & $9.19\times10^{-5}$ (Al) 
&  ($1.80\times10^{-3}$) & ($1.64\times10^{-3}$)
& ($2.52\times10^{-4}$)&
($2.51\times10^{-4}$)\\
\hline
\multirow{3}*{$|\lambda_{233}\lambda_{133}^{*}|$}
& $1.53\times10^{-3}$ (Ti) & $1.53\times10^{-3}$ (Ti) & & & &\\
& $1.74\times10^{-3}$ (Au) & $1.73\times10^{-3}$ (Au) 
&  $2.85\times10^{-3}$ & $2.59\times10^{-3}$
&  $8.45\times10^{-3}$ & 
$8.44\times10^{-3}$\\
& $1.38\times10^{-4}$ (Al) & $1.38\times10^{-4}$ (Al) 
&  ($1.80\times10^{-3}$) & ($1.64\times10^{-3}$)
& ($3.77\times10^{-4}$)& ($3.77\times10^{-4}$)\\
\hline\hline
\end{tabular}}
\caption{The upper limits with and without RG effects on the combinations of $\lambda$ couplings from cLFV processes $\mu$-$e$ conversion, $\mu\to e\gamma $, and $\mu\to 3 e$. The units of the upper limits are $[m_{\tilde{\ell}}^{2} ~(\text{TeV}^{-2})]$.}\label{lambda-results}
\end{table}

In Tab.~\ref{lambda-results}, the upper limits on the combinations of $\lambda$ have been shown. As the first three combinations can give tree-level contributions to $\mu\to 3e$ decay, while giving one-loop contributions to $\mu$-$e$ conversion and $\mu\to e\gamma$, the constraints from $\mu\to 3e$ are stronger than those from the other two processes. The RG effects can improve the upper limits by no more than 5\%. In the last three cases, the contributions to the three cLFV processes are all at the one-loop level. The future $\mu$-$e$ conversion experiment can give stronger limits than the other two processes. 

The neutrino mass can also impose stringent limits on combinations of $\lambda$. The neutrino mass can be generated similarly to Eq.~(\ref{vmass-1}) by replacing the quark mass with lepton mass, using  coefficients $1/8\pi^2$, and defining $\tilde{m}=m_{\tilde{\ell}}^{2}/(A^{e}-\mu\tan\beta)$.
The combination $\lambda_{233}\lambda_{133}$ is dominant and can be constrained to be $<3\times 10^{-6}$ with $\tilde{m}$ at the TeV scale. The combination $\lambda_{121}\lambda_{122}$ can also contribute to the neutrino mass, but is suppressed by $m_{e}m_{\mu}/\tilde{m}$.

\section{Conclusion and Prospects}
The $\mu$–$e$ conversion process provides a powerful and highly sensitive means to probe charged lepton flavor violation. This work revisits the muon charged lepton flavor violation (cLFV) processes in the Supersymmetric model with trilinear $R$-parity violation to figure out the ability of future $\mu$-$e$ conversion experiments on examining the parameter space. 

The model is treated within the effective field theory framework, following the standard procedure for matching and running. We present certain matching results for the SUSY model onto SMEFT operators at tree and one-loop levels at the new physics scale, relevant to cLFV processes. The matching conditions from SMEFT to LEFT at the electroweak scale with the RG running effects are numerically concluded.

The constraints on the combinations of coupling parameters in the form of $|\lambda^{(\prime)}_{ijk}\lambda^{(\prime)*}_{prs}|$ with units given as $[m_{\tilde{f}}^2(\text{TeV}^{-2})]$ are derived. To simplify the discussion, we assume that only one of those combinations is nonzero at a time. There are 15 combinations of $\lambda^{\prime}$ and 6 combinations of $\lambda$ that have been investigated, some of which are lacking in the previous literature, which focuses on muon cLFV within the $R$-parity trilinear violating SUSY model. The constraints on the combinations from neutrinoless double beta decay, neutrino mass, and certain meson decay processes are also discussed. 

\begin{figure}[t]
\centering
\includegraphics[width=0.95\textwidth]{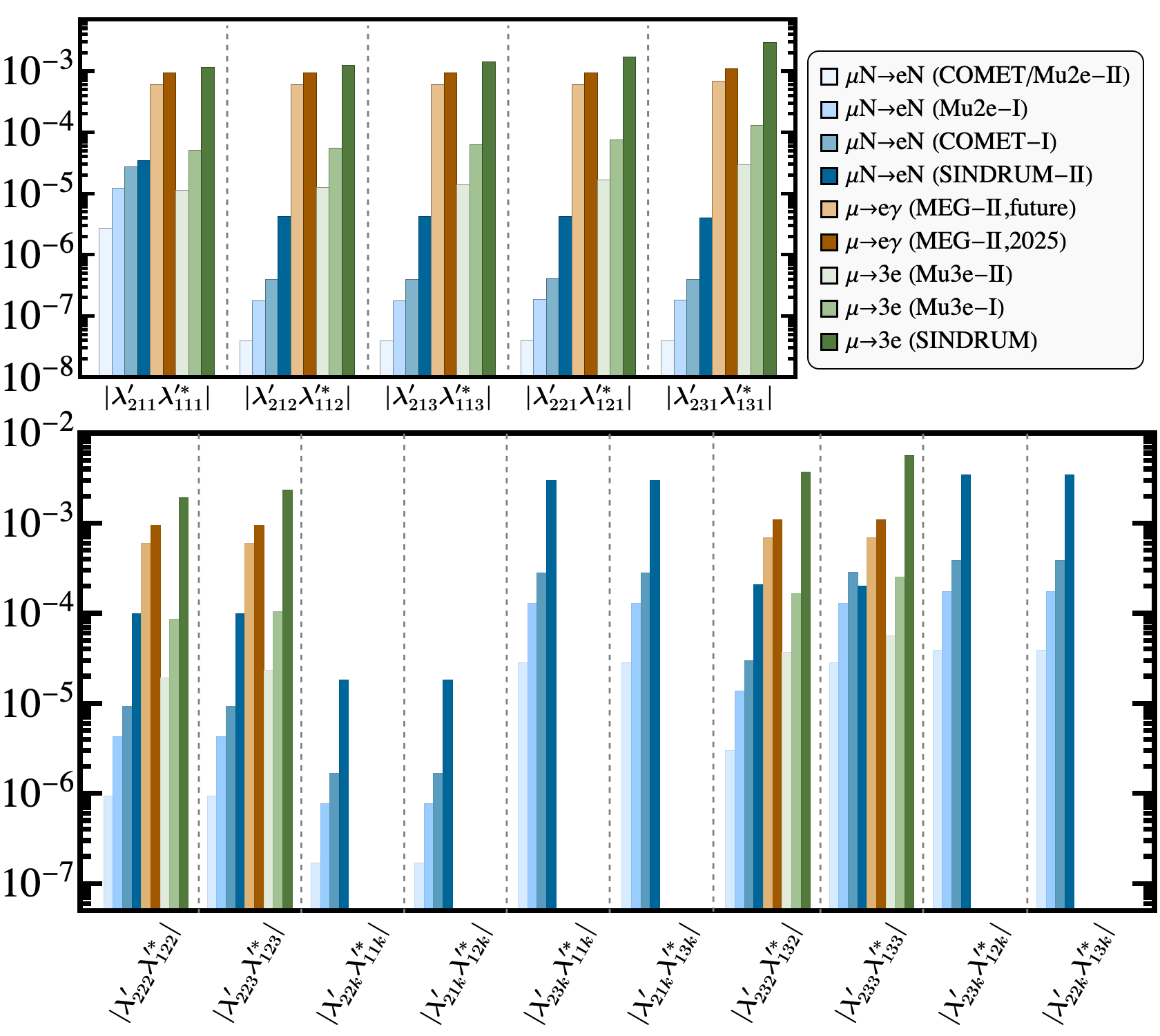}
\caption{The upper limits with RG effects on the combinations of $\lambda^{\prime}$ couplings from muon cLFV processes $\mu$-$e$ conversion (in blue), $\mu\to e\gamma $ (in orange), and $\mu\to 3 e$ (in green). The units of the upper limits are $[m_{\tilde{q}}^{2} ~(\text{TeV}^{-2})]$.}\label{Fig1}
\end{figure}
\begin{figure}[t]
\centering
\includegraphics[width=0.95\textwidth]{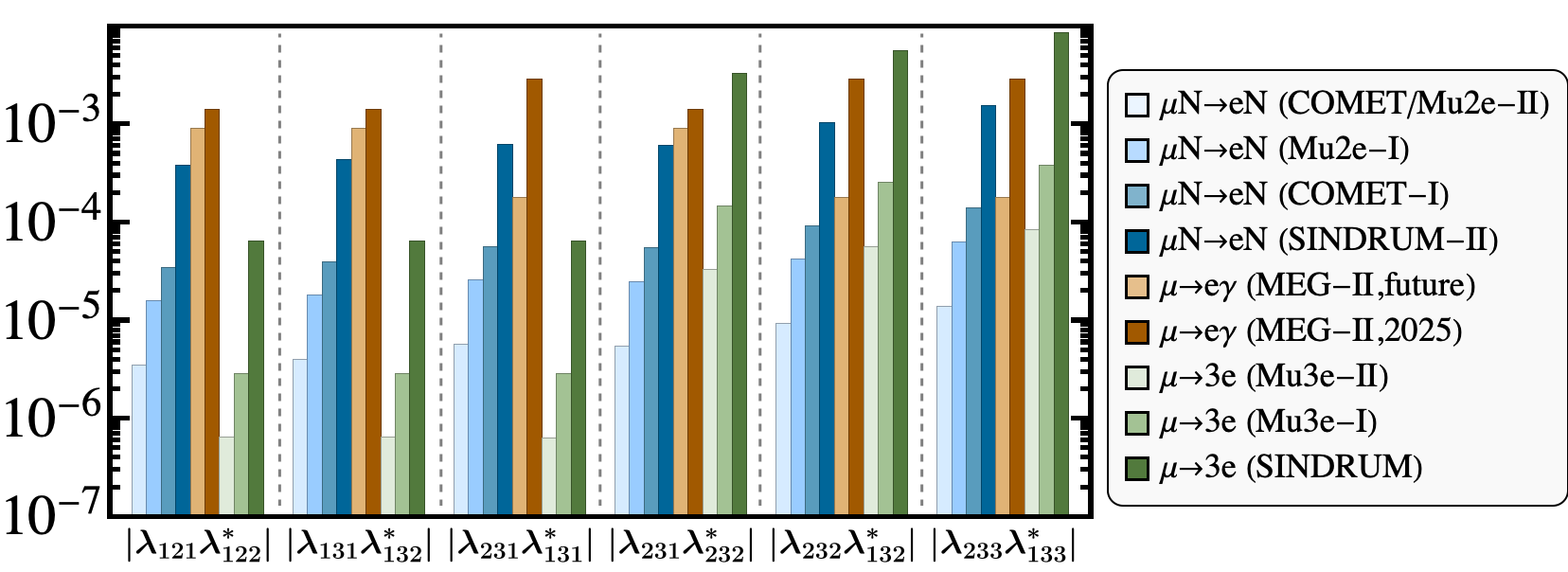}
\caption{The upper limits with RG effects on the combinations of $\lambda$ couplings from muon cLFV processes $\mu$-$e$ conversion (in blue), $\mu\to e\gamma $ (in orange), and $\mu\to 3 e$ (in green). The units of the upper limits are $[m_{\tilde{\ell}}^{2} ~(\text{TeV}^{-2})]$.}\label{Fig2}
\end{figure}

This paper give new upper limits for trilinear parameters, constrained by the latest experimental data and future sensitivities. It also qualitatively highlights the importance of RG effects. The upper limits on the parameters of $\lambda^{\prime}$ term are shown in Tab.~\ref{lambdap-results} and Tab.~\ref{lambdap-results2}, while the limits on $\lambda$ parameters are shown in Tab.~\ref{lambda-results}. These upper limits are also shown in Fig.~\ref{Fig1} and Fig.~\ref{Fig2} to give a more intuitive view, which also include the sensitivities of Mu2e Run-II, COMET phase II, and Mu3e phase II for a comprehensive comparison. One can conclude that, for most combinations, the future $\mu$-$e$ conversion experiments, COMET phase I, which are expected to begin data-taking in the near future, will yield stronger limits than the other two $\mu\to e\gamma$ and $\mu\to 3e$ decay processes. In certain cases, $\mu$-$e$ conversion is the only process that can constrain the combinations, as the other two processes are GIM suppressed. We can also conclude that the RG effects can play an important role in certain cases. The RG effects can give no more than 30\% influence on the results in most of the combinations, while can give 80\% improvement on the upper limits for $\lambda_{23k}^{\prime} \lambda_{12k}^{\prime *}$ and $\lambda_{13k}^{\prime} \lambda_{22k}^{\prime *}$ combinations.

Taking future experimental scenarios into account, we can make the following prospects. 
If no signals are observed in future experiments, the parameters will be constrained. If one cLFV experiment detects a signal, the other two can validate the parameter space. For example, if a signal appears in $\mu$-$e$ conversion but not in $\mu\to 3e$, this would very likely exclude the first three $\lambda$ combinations in Tab.~\ref{lambda-results}. Conversely, if $\mu$-$e$ conversion is never observed in COMET phase I but $\mu\to 3e$ or $\mu\to e\gamma$ are in MEG-II or Mu3e phase I, which sensitivities are shown in Tab.~\ref{exp}, this would suggest that most combinations could be ruled out.
The combinations $\lambda_{133}^{\prime}\lambda_{233}^{\prime}$ and $\lambda_{133}\lambda_{233}$ are also constrained by neutrino mass, whose limits are much stronger than near future cLFV experiments can give. This can reveal that if the neutrino masses can be dominantly realized by those combinations, the contributions to cLFV processes will be strongly suppressed and not observable in the near future cLFV experimental searches. These prospects assume a single nonzero combination. The multiple nonzero combinations may lead to different results requiring further analysis.

In conclusion, the next-generation $\mu$–$e$ conversion searches will be indispensable for probing new physics. With the careful treatment of the RG effect, $\mu$-$e$ conversion experiments are expected to offer superior sensitivity and give more comprehensive examinations of the models compared to $\mu\to e\gamma$ and $\mu\to 3 e$ in specific models and cases. Future experiments will be beneficial for a deeper understanding of lepton flavor violation mechanisms and the underlying new physics contributions.

\section*{Acknowledgements}
Y.-Q. Xiao thanks Shao-Long Chen, Xiao-Dong Ma, Michael J. Ramsey-Musolf and Xiang Zhao for the useful discussion and comments. X.-G. He is supported in part by NSFC (Nos. 12090064, 12375088, W2441004). Y.-Q. Xiao is partly supported by NSFC (No. 12547133).

\bibliographystyle{JHEP}
\bibliography{references}

\end{document}